\documentclass[preprint,secnumarabic,amssymb,nobibnotes,nofootinbib,aps,prc]{revtex4}
\usepackage{graphicx}
\usepackage{bm}
\newcommand{\bea}{\begin{eqnarray}}
\newcommand{\eea}{\end{eqnarray}}
\newcommand{\be}{\begin{equation}}
\newcommand{\ee}{\end{equation}}
\newcommand{\bt}{\begin{tabular}}
\newcommand{\et}{\end{tabular}}
\newcommand{\Tr}{{\rm Tr}}
\newcommand{\no}{\nonumber}
\newcommand{\ovl}{\overline}

\newcommand{\Si}{\Sigma}

\newcommand{\pa}{\partial}
\newcommand{\pam}{\partial_\mu}
\newcommand{\beas}{\begin{eqnarray*}}
\newcommand{\eeas}{\end{eqnarray*}}

\newcommand{\fr}{\frac}

\newcommand{\dg}{\dagger}
\newcommand{\La}{\Lambda}
\begin{document}
\baselineskip 6.5mm
\date{\today}
\title{Effects of Dirac sea polarization on hadronic properties\\
        -A chiral SU(3) approach}
%%%%%%%%%%%%%%%%%%%%%%%%%%%%%%%%%%%%%%%%%%%%%%%%%%%%%%%%%%%
\author{A.Mishra}
\email{mishra@th.physik.uni-frankfurt.de}
\affiliation{Institut f\"ur Theoretische Physik,
        Robert Mayer Str. 8-10, D-60054 Frankfurt am Main, Germany}
\author{K.Balazs}
\email{balazs@th.physik.uni-frankfurt.de}
\affiliation{Institut f\"ur Theoretische Physik,
        Robert Mayer Str. 8-10, D-60054 Frankfurt am Main, Germany}
\author{D.Zschiesche}
\email{ziesche@th.physik.uni-frankfurt.de}
\affiliation{Institut f\"ur Theoretische Physik,
        Robert Mayer Str. 8-10, D-60054 Frankfurt am Main, Germany}
\author{S. Schramm}
\affiliation{Institut f\"ur Theoretische Physik,
        Robert Mayer Str. 8-10, D-60054 Frankfurt am Main, Germany}
\author{H.~St\"ocker}
\affiliation{Institut f\"ur Theoretische Physik,
        Robert Mayer Str. 8-10, D-60054 Frankfurt am Main, Germany}
\author{W.~Greiner}
\affiliation{Institut f\"ur Theoretische Physik,
       Robert Mayer Str. 8-10, D-60054 Frankfurt am Main, Germany}
%%%%%%%%%%%%%%%%%%%%%%%%%%%%%%%%%%%%%%%%%%%%%%%%%%%%%%%%%%%%%
\begin{abstract}
The effect of vacuum fluctuations on the in-medium hadronic
properties is investigated using a chiral SU(3) model in the
nonlinear realization. The effect of the baryon Dirac sea is seen
to modify hadronic properties and in contrast to a calculation 
in mean field approximation it is seen to give rise to a significant 
drop of the vector meson masses in hot and dense matter. This effect 
is taken into account through the summation of baryonic tadpole 
diagrams in the relativistic Hartree approximation (RHA),
where the baryon self energy is modified due to interactions with
both the non-strange $(\sigma)$ and the strange $(\zeta)$ scalar
fields.
\end{abstract}
%\pacs{}
\maketitle
\def\bfm#1{\mbox{\boldmath $#1$}}
%
%%%%%%%%%%%%%%%%%%%%%%%%%%%%%%%%%%%%%%%%%%%%%%%%%%%%%%%%%%%%%
%
\section{ Introduction }
The study of hot and dense matter is an important problem in strong
interaction physics. In recent times there have been numerous experimental
investigations , e.g. in the context of relativistic heavy ion collision
experiments, to study how hadronic matter is modified under extreme
conditions of high temperatures and/or densities
\cite{helios,ceres,dls,rhic,hades,brown,rapp,hat,jin,samir,
weise,pisa,csong,mharada,ernst}.
The experimental observables from relativistic heavy ion collision
are related to the medium modifications of the hadrons
in the dynamically evolving strongly interacting matter (fireball)
resulting from the nuclear collision. One of the explanations of the
observed
enhanced dilepton production in the low invariant mass regime
is the medium modification of the vector mesons \cite{helios,ceres,dls}.
It was first conjectured to be a simple scaling law for the
vector meson masses in the medium \cite{brown}. There have been also
QCD sum rule calculations for studying the in-medium vector meson
properties  \cite{hat,jin,samir,weise}.
In the Quantum Hadrodynamics framework \cite{qhd}, the vector meson
masses were shown to have dominant contributions from the nucleon Dirac
sea. The vector meson masses have very insignificant modifications in
the hot and dense medium, when the contributions only from the
Fermi sea are taken into account  \cite{hatsuda,hatsuda1,jeans,sourav,caillon}.
Recently, it was shown in a chiral SU(3) model \cite{paper3,springer},
that the Dirac sea polarization leads to a significant drop of vector meson
masses in nuclear matter \cite{hartree}. In the present investigation, we
study the properties of hadrons in the hot hyperonic matter, taking
into account the effects of the Dirac sea polarization
\cite{vecmass,dlp,mishra}.\\

We organize the paper as follows. In the section 2, we briefly
recapitulate the SU(3) chiral model used in the present investigation.
In section 3, we outline the mean field approximation for the study
of hadronic properties. Section 4 discusses the inclusion of vacuum
polarization effects using RHA. This is done by summing over the baryonic
tadpole diagrams which includes couplings to both the nonstrange
($\sigma$) and strange ($\zeta$) scalar meson fields. In section 5 
we discuss the medium modification of the vector meson masses due to their
interaction with the baryons in the hot hadronic matter.
The effect of vacuum polarization compared to the mean field approximation
is studied. We discuss the results of the present investigation in
section 6. Finally, in section 7, we summarize our findings and discuss
possible improvements of the current approach.
%%%%%%%%%%%%%%%%%%%%%%%%%%%%%%%%%%%%%%%%%%%%%%%%%%%%%%%%%%%%%%%%

\section{ The hadronic chiral $SU(3) \times SU(3)$ model }
\label{model}
In this section the various terms of the Lagrangian
\be
\label{geslag}
{\cal L} = {\cal L}_{kin} + \sum_{ W =X,Y,V,{\cal A},u }{\cal L}_{BW}
          + {\cal L}_{VP} + {\cal L}_{vec} + {\cal L}_0 + {\cal L}_{SB}
\ee
are discussed. The calculation is done within the framework of
a relativistic quantum field theoretical
model of baryons and mesons built on chiral symmetry and broken scale
invariance \cite{paper3} to describe strongly interacting nuclear matter.
We adopt a nonlinear realization of the chiral symmetry which allows a
simultaneous description of hyperon potentials and properties of
finite nuclei \cite{deformed,paper3,springer}.
This Lagrangian contains the baryon octet, the spin-0 and spin-1 meson
multiplets as degrees of freedom.\\
$ {\cal L}_{kin} $ is the kinetic energy term, $  {\cal L}_{BW}  $ contains the
baryon-meson interactions in which the baryon-spin-0 meson interaction
terms generate the baryon masses. $ {\cal L}_{VP} $ describes the interactions
of vector mesons with the pseudoscalar mesons (and with photons).
$ {\cal L}_{vec} $ describes the dynamical mass generation of the vector
mesons through coupling to the scalar fields and contains
additionally quartic self-interactions of the vector fields.
$ {\cal L}_0 $ are the meson-meson interaction terms inducing the
spontaneous breaking of chiral symmetry. It also includes
a scale invariance breaking logarithmic potential. $ {\cal L}_{SB} $
describes the explicit symmetry breaking of $ U(1)_A $, $ SU(3)_V $
and the chiral symmetry.
\subsection{ The kinetic energy terms }
\label{kinterm}
An important property of the nonlinear realization of chiral symmetry
is that all terms of the model-Lagrangian only have to be invariant
under the $SU(3)_V$ transformation in order to ensure chiral
symmetry. This vector transformation depends in general on the
pseudoscalar mesons and thus is local. Covariant derivatives have
to be introduced for the kinetic terms in order to preserve chiral
invariance \cite{paper3}. The covariant derivative used in this case, reads:
$ D_\mu = \pam\, + [\Gamma_\mu,\,\,]$
with $\Gamma_\mu=-\fr{i}{2}[u^\dg\pam u + u\pam u^\dg]$ where
$u=\exp\Bigg[\fr{i}{\sigma_0}\pi^a\lambda^a\gamma_5\Bigg]$ is the unitary
transformation operator  \cite{paper3}. The pseudoscalar mesons are given as parameters
of the symmetry transformation. \\
In summary, the kinetic energy terms read \cite{paper3}
\bea
\label{kinlag}
{\cal L}_{kin} &=& i\Tr \overline{B} \gamma_{\mu} D^{\mu}B
                + \frac{1}{2} \Tr D_{\mu} X D^{\mu} X
+  \Tr (u_{\mu} X u^{\mu}X +X u_{\mu} u^{\mu} X)
                + \frac{1}{2}\Tr D_{\mu} Y D^{\mu} Y \nonumber \\
               &+&\frac {1}{2} D_{\mu} \chi D^{\mu} \chi
                - \frac{ 1 }{ 4 } \Tr
\left(\tilde V_{ \mu \nu } \tilde V^{\mu \nu }  \right)
- \frac{ 1 }{ 4 } \Tr \left(F_{ \mu \nu } F^{\mu \nu }  \right)
- \frac{ 1 }{ 4 } \Tr \left( {\cal A}_{ \mu \nu } {\cal A}^{\mu \nu }  \right)\, .
\eea
B denotes the baryon octet, X the scalar meson multiplet, Y the pseudoscalar
chiral singlet, $\tilde{V}^\mu$ (${\cal A}^\mu$) the renormalised vector
(axial vector) meson multiplet with the field strength tensor
$\tilde{V}_{\mu\nu}=\pa_\mu\tilde{V}_\nu-\pa_\nu\tilde{V}_\mu$
$({\cal A}_{\mu\nu}= \pa_\mu{\cal A}_\nu-\pa_\nu{\cal A}_\mu $).
$F_{\mu\nu}$ is the electro-magnetic field tensor and
$\chi$ is the scalar, iso-scalar dilaton (glueball) -field.
\subsection{Baryon-meson interaction}
\label{bmwwterm}
Except for the difference in Lorentz indices, the SU(3) structure of the
spin-1/2 baryon -meson interaction terms are the same for all mesons.
This interaction for a general meson field $W$ has the form
\be
\label{bmwwlag}
{\cal L}_{BW} =
-\sqrt{2}g_8^W \left(\alpha_W[\ovl{B}{\cal O}BW]_F+ (1-\alpha_W)
[\ovl{B} {\cal O}B W]_D \right)
- g_1^W \frac{1}{\sqrt{3}} \Tr(\ovl{B}{\cal O} B)\Tr W  \, ,
\ee
with $[\ovl{B}{\cal O}BW]_F:=\Tr(\ovl{B}{\cal O}WB-\ovl{B}{\cal O}BW)$ and
$[\ovl{B}{\cal O}BW]_D:= \Tr(\ovl{B}{\cal O}WB+\ovl{B}{\cal O}BW)
- \frac{2}{3}\Tr (\ovl{B}{\cal O} B) \Tr W$.
The different terms to be considered are those for the interaction
of baryons  with
scalar mesons ($W=X, {\cal O}=1$), with
vector mesons  ($W=\tilde V_{\mu}, {\cal O}=\gamma_{\mu}$ for the vector and
$W=\tilde V_{\mu \nu}, {\cal O}=\sigma^{\mu \nu}$ for the tensor
interaction),
with axial vector mesons ($W={\cal A}_\mu, {\cal O}=\gamma_\mu \gamma_5$)
and with
pseudoscalar mesons ($W=u_{\mu},{\cal O}=\gamma_{\mu}\gamma_5$), respectively.
For the current investigation the following interactions are relevant.
\subsubsection{\underline{Baryon-scalar meson interaction}}
\label{bswwterm}
This is the term generating the baryon masses through coupling of the baryons
to the non-strange $ \sigma (\sim \langle\bar{u}u + \bar{d}d\rangle) $
and the strange $ \zeta(\sim\langle\bar{s}s\rangle) $ scalar quark condensate
\cite{paper3}.
After insertion of the scalar meson matrix $X$, one obtains the baryon masses
\bea
\label{bmassen}
m_N &=& m_0 - \fr{1}{3}g^S_8(4\alpha_S-1)(\sqrt{2}\zeta - \sigma),
  \nonumber  \\
m_{\Lambda} &=& m_0 - \fr{2}{3}g^S_8(\alpha_S-1)(\sqrt{2}\zeta - \sigma),
  \nonumber  \\
m_{\Sigma} &=& m_0 + \fr{2}{3}g^S_8(\alpha_S-1)(\sqrt{2}\zeta - \sigma),
 \\
m_{\Xi} &=& m_0 + \fr{1}{3}g^S_8(2\alpha_S + 1)(\sqrt{2}\zeta - \sigma),
  \nonumber
\eea
with $m_0=g^S_1(\sqrt{2}\sigma + \zeta)/\sqrt{3}$. The parameters $ g_1^S , g_8^S $
and $ \alpha_S $ can be used to fix the baryon masses to their experimentally measured
vacuum values. It should be emphasized that the nucleon mass also depends on the
{\em strange condensate} $ \zeta $. This general case will be used in the present
investigation, to study hot and strange hadronic matter. Recently the vector meson
masses were investigated in nuclear matter\cite{hartree}, for the situation 
$ \alpha_S = 1 $ and $ g_1^S = \sqrt{6}g_8^S $, where the nucleon mass 
depends only on the non-strange quark condensate \cite{hartree,paper3}.\\
The effect of including RHA is similar to the results obtained
in the Walecka model \cite{hatsuda,vecmass}.
In the present investigation, however, in the summing over baryon
tadpoles the effects of coupling of baryons to both scalar fields
($\sigma$ and $\zeta$) have to be taken into account for the baryonic
Dirac sea in RHA.
\subsubsection{\underline{Baryon-vector meson interaction}}
\label{bvwwterm}
In analogy to the baryon-scalar meson coupling there exist two independent baryon-vector
meson interaction terms corresponding to the F-type (antisymmetric) and D-type
(symmetric) couplings. Here we will use the symmetric coupling because from
the universality principle  \cite{saku69} and the vector meson dominance model one can conclude
that the antisymmetric coupling should be small.
We realize this assumption by setting $\alpha_V=1$
for all fits. Additionally we decouple the strange vector field
$ \phi_\mu\sim\bar{s} \gamma_\mu s $ from the nucleon by setting
$ g_1^V=\sqrt{6}g_8^V $ and the remaining
baryon-vector meson interaction reads
\be
\label{bvwwlag}
{\cal L}_{BV}=-\sqrt{2}g_8^V\Big\{[\bar{B}\gamma_\mu BV^\mu]_F+\Tr\big(\bar{B}\gamma_\mu B\big)
\Tr V^\mu\Big\}\, .
\ee
Note that in this limit all coupling constants are fixed once $g_8^V$ is
specified \cite{paper3}.
This is done by fitting the nucleon-$\omega$ coupling
to the energy density at nuclear
matter saturation ($E/A = -16$~MeV).
With the above choice, the vector meson- baryon couplings reduce to those
from the additive quark model given as
\bea
g_{\Lambda \omega} &= & \frac{2}{3} g_{N\omega}=
g_{\Sigma \omega}=2 g_{ \Xi \omega}
\nonumber  \\
g_{\Lambda \phi} &= & -\frac{\sqrt 2}{3} g_{N\omega}=
g_{\Sigma \phi}=2 g_{ \Xi \phi}
\label{gvb}
\eea
\subsection{Meson-meson interaction}
\label{mmwwterm}
%
%%%%%%%%%%%%%%%corrected upto here%%%%%%%%%%%%%%%%%%%%%%%%%
\subsubsection{\underline{Spin-0 potential}}
\label{chipot}
The Lagrangian describing the interaction for the scalar mesons, $X$,
and pseudoscalar singlet, $Y$, is given as \cite{paper3}
\bea
\label{chipotlag}
{\cal L}_0 &= &  -\frac{ 1 }{ 2 } k_0 \chi^2 I_2
     + k_1 (I_2)^2 + k_2 I_4 +2 k_3 \chi I_3,
\eea
with $I_2= \Tr (X+iY)^2$, $I_3=\det (X+iY)$ and $I_4 = \Tr (X+iY)^4$.
The scalar glueball field $\chi$ is
introduced to satisfy the QCD trace anomaly i.e. nonvanishing
energy-momentum tensor $\Theta_\mu^\mu = (\beta_{QCD}/2g)\langle
G^a_{\mu\nu}G^{a,\mu\nu}\rangle$, where $G^a_{\mu\nu}$ is the gluon
field tensor.  \\
A scale breaking potential \cite{sche80}
\be
\label{scalelag}
  {\cal L}_{\mathrm{scalebreak}}=- \frac{1}{4}\chi^4 \ln \frac{ \chi^4 }{ \chi_0^4}
 +\frac{\delta}{3}\chi^4 \ln \frac{I_3}{\det \langle X \rangle_0}
\ee
is introduced. This yields
\be
\theta_\mu^\mu=4{\cal L}-\chi\fr{\pa{\cal L}}{\pa\chi}-2\pa_\mu\chi
\fr{\pa{\cal L}}{\pa(\pa_\mu\chi)} = \chi^4
\ee
and allows for the identification of the $\chi$ field width the gluon condensate
$\Theta_\mu^\mu=(1-\delta)\theta_\mu^\mu=(1-\delta)\chi^4$.
Finally the term
%\be
%\label{quartic}
${\cal L}_{\chi} = - k_4 \chi^4$
%\ee
generates a phenomenologically consistent finite vacuum expectation
value. We shall use the frozen glueball approximation i.e. assume
$\chi = \langle 0|\chi|0\rangle\equiv\chi_0$, since the variation of
$\chi$ in the medium is rather small \cite{paper3}.
\subsubsection{\underline{Vector mesons masses}}
\label{vecmassen}
The Lagrangian for the vector meson interaction is written as \cite{gasi69,mitt68,toki}
\bea
\label{veclag}
{\cal L}_{vec} &=&
    \fr{1}{2}m_V^2\fr{\chi^2}{\chi_0^2}\Tr\big(\tilde{V}_\mu\tilde{V}^\mu\big)
+   \fr{1}{4}\mu\Tr\big(\tilde{V}_{\mu\nu}\tilde{V}^{\mu\nu}X^2\big)\\ \no
&+& \fr{1}{12}\lambda_V\Big(\Tr\big(\tilde{V}^{\mu\nu}\big)\Big)^2 +
    2(\tilde{g}_4)^4\Tr\big(\tilde{V}_\mu\tilde{V}^\mu\big)^2  \, .
\eea
The vector meson fields, $\tilde{V}_\mu$ are related to the
renormalized fields by
$
%\label{renvfields}
V_\mu = Z_V^{1/2}\tilde{V}_\mu \quad\mbox{with}\quad V = \omega, \rho, \phi \,
$ \cite{hartree}.
The masses of $\omega,\rho$ and $\phi$ are fitted by tuning $m_V$, $\mu$ and
$\lambda_V$. The vector meson masses have contributions from
the quartic self-interaction,
and we get in the frozen glueball approximation
\bea
\label{vmass}
{m_\omega^\ast}^2 &=& m_\omega^2 + 12g_4^4\omega^2 \, ,\nonumber \\
{m_\rho^\ast}^2 &=& m_\rho^2+12g_4^4\fr{Z_\rho}{Z_\omega}\omega^2 \, , \\
{m_\phi^\ast}^2 &=& m_\phi^2+24g_4^4\fr{Z_\phi^2}{Z_\omega^2}\phi^2 \, ,\nonumber
\eea
with $g_4 = \sqrt{Z_\omega} \tilde g_4$ as the renormalized coupling.
Since the quartic vector-interaction contributes only in the medium,
the coupling $g_4$ cannot be unambiguously fixed.
It is fitted, so that the compressibility lies
in the desired region between $200-300 \mbox{ MeV}$ in the mean field
approximation. Note that the $N-\omega$ as well as the
$N-\rho$ - couplings are also affected by the redefinition of the
fields with the corresponding renormalised coupling constants as
$g_{N\omega} \equiv 3 g_V^8 \sqrt{Z_\omega}$ and
$g_{N\rho} \equiv g_V^8 \sqrt{Z_\rho}$.
\subsection{Explicit chiral symmetry breaking}
The explicit symmetry breaking term is given as \cite{paper3}
\be
\label{esblag}
 {\cal L}_{SB}=\Tr A_p\left(u(X+iY)u+u^\dagger(X-iY)u^\dagger\right)
\ee
with $A_p=1/\sqrt{2}{\mathrm{diag}}(m_{\pi}^2 f_{\pi},m_\pi^2 f_\pi, 2 m_K^2 f_K
-m_{\pi}^2 f_\pi)$ and $m_{\pi}=139$ MeV, $m_K=498$ MeV. This
choice for $A_p$ together with the constraints
%\be
%\label{scalar0}
$
\sigma_0 = -f_{\pi}\;, \zeta_0 = -\frac{1}{\sqrt{2}}(2 f_K - f_{\pi}),
$
%\ee
for the VEV on the scalar condensates assure that
the PCAC-relations of the pion and kaon are fulfilled.
With $f_{\pi} = 93.3$~MeV and $f_K = 122$~MeV we obtain $|\sigma_0| =
93.3$~MeV and $|\zeta_0| = 106.56$~MeV.
\section{The mean field approximation}
\label{mft}
The Lagrangian density in the mean field approximation \cite{paper3}
consists of the following terms
\begin{eqnarray}
\label{mftlag}
{\cal L}_{BX}+{\cal L}_{BV} &=& -\sum_i\overline{\psi_{i}}\, [g_{i
\omega}\gamma_0 \omega + g_{i\phi}\gamma_0 \phi
+m_i^{\ast} ]\,\psi_{i} \\
{\cal L}_{vec} &=& \frac{1}{2}m_{\omega}^{2}\frac{\chi^2}{\chi_0^2}\omega^
2+g_4^4 \omega^4 +
\frac{1}{2}m_{\phi}^{2}\frac{\chi^2}{\chi_0^2}\phi^2+g_4^4
\left(\fr{Z_\phi}{Z_\omega}\right)^2\phi^4\\
{\cal V}_0 &=& \frac{ 1 }{ 2 } k_0 \chi^2 (\sigma^2+\zeta^2)
- k_1 (\sigma^2+\zeta^2)^2
     - k_2 ( \frac{ \sigma^4}{ 2 } + \zeta^4)
     - k_3 \chi \sigma^2 \zeta \nonumber \\
&+& k_4 \chi^4 + \frac{1}{4}\chi^4 \ln \frac{ \chi^4 }{ \chi_0^4}
 -\frac{\delta}{3} \chi^4 \ln \frac{\sigma^2\zeta}{\sigma_0^2 \zeta_0} \\
{\cal V}_{SB} &=& \left(\frac{\chi}{\chi_0}\right)^{2}\left[m_{\pi}^2 f_{\pi}
\sigma
+ (\sqrt{2}m_K^2 f_K - \frac{ 1 }{ \sqrt{2} } m_{\pi}^2 f_{\pi})\zeta
\right],
\end{eqnarray}
where $m_i^* = -g_{\sigma i}{\sigma}-g_{\zeta i}{\zeta} $ is the
effective mass of the baryon of type $i$, with $i = N,\Si ,\La ,\Xi$.
The thermodynamical potential of the grand
canonical ensemble, $\Omega$, per unit volume $V$ at given chemical
potential $\mu$ and temperature $T$ can be written as
\bea
\label{thermpot}
\frac{\Omega}{V} &=& -{\cal L}_{vec} - {\cal L}_0 - {\cal L}_{SB}
- {\cal V}_{vac} + \sum_i\frac{\gamma_i }{(2 \pi)^3}
\int d^3k\,
E^{\ast}_i(k)\Big(f_i(k)+\bar{f}_i(k)
\Big) \nonumber \\
&&- \sum_i\frac{\gamma_i }{(2 \pi)^3}\,\mu^{\ast}_i
\int d^3k\,\Big(f_i(k)-\bar{f}_i(k)\Big)\, .
\eea
Here the vacuum energy (the potential at $\rho=0$) has been substracted 
in order to obtain a vanishing vacuum energy. $\gamma_i$ is the 
spin-isospin degeneracy factor for baryon, $i$ and 
$E^{\ast}_i(k) = \sqrt{k_i^2+{m^\ast_i}^2}$ and 
$ \mu^{\ast}_i = \mu_i-g_{i\omega}\omega$ are the effective single particle
energy and effective chemical potential respectively.
$f_i$ and $\bar{f}_i$ are the thermal distribution functions for
the baryons and antibaryons given as 
\be
f_i(k) = \fr{1}{{\rm e}^{\beta (E^{\ast}_i(k)-\mu^{\ast}_i)}+1}\quad ,\quad
\bar{f}_i(k)=\fr{1}{{\rm e}^{\beta (E^{\ast}_i(k)+\mu^{\ast}_i)}+1}. \no 
\ee
The mesonic field equations are determined by minimizing the
thermodynamical potential
\bea
\label{mesoneom}
\frac{\partial (\Omega/V)}{\partial \sigma}\Bigg|_{MFT} &=&
  k_0 \chi^2 \sigma - 4 k_1 (\sigma^2+\zeta^2)\sigma
- 2k_2 \sigma^3        - 2 k_3 \chi \sigma \zeta
- 2\frac{\delta \chi^4}{3 \sigma} + \nonumber \\
&+& m_{\pi}^2 f_{\pi}
+ \sum_i\frac{\pa m_i^{\ast}}{\pa \sigma}\rho^s_i=0 \, , \\
\frac{\partial (\Omega/V)}{\partial \zeta}\Bigg|_{MFT} &=&
  k_0 \chi^2 \zeta - 4 k_1 (\sigma^2+\zeta^2) \zeta
- 4 k_2 \zeta ^3 - k_3 \chi \sigma^2
- \frac{\delta \chi^4}{3 \zeta} +\nonumber \\
&+& \left[\sqrt{2}m_K^2 f_K
- \frac{ 1 }{ \sqrt{2}} m_{\pi}^2 f_{\pi}\right]
+ \sum_i\frac{\pa m_i^{\ast}}{\pa \zeta}\rho^s_i=0 \, , \\
\frac{\partial (\Omega/V)}{\partial \omega}\Bigg|_{MFT} &=&
-m_{\omega}^2 \omega - 4 g_4^4 \omega^3 + \sum_i g_{i \omega}\rho_i =
0 \, , \\
\frac{\partial (\Omega/V)}{\partial \phi}\Bigg|_{MFT} &=&
-m_{\phi}^2 \phi - 8 g_4^4\left(\fr{Z_\phi}{Z_\omega}\right)^2
\phi^3 + \sum_i g_{i \phi}\rho_i = 0 \, .
\eea
In the above, $\rho^s_i$ and $\rho_i$ are the scalar and vector densities
for the baryons at finite temperature given as,
\bea
\label{dichten}
\rho^s_i = \gamma_i
\int \frac{d^3 k}{(2 \pi)^3} \,\frac{m_i^{\ast}}{E^{\ast}_i}\,
\left(f_i(k) + \bar{f}_i(k)\right) \, , \; \; \; \;
\rho_i = \gamma_i \int \frac{d^3 k}{(2 \pi)^3}\,\left(f_i(k) -
\bar{f}_i(k)\right) \,.
\eea
The energy density and the pressure follow from the Gibbs-Duhem relation,
$\epsilon = \Omega/V+\mu_i\rho_i+TS $ and $ p = -\Omega/V $.
%%%%%%%%%%%%%%%%%%%%%%%%%%%%%%%%%%%%%%%%%%%%%%%%%%%%%%%%%%%%
\section{The relativistic Hartree approximation}
\label{rha}
The relativistic Hartree approximation takes into account the effects from
the Dirac sea through evaluating the baryonic tadpole diagrams. The
interacting propagator for baryon of type $i$ is given by the Schwinger-Dyson
equation
\be
\label{sdeqn-bar}
G_i^H(p) = G_i^0(p)+G_i^0(p)\Sigma_iG_i^H(p)\, ,
\ee
with $G_i^0(p)$ as the free propagator and $\Sigma_i(p)$ as the
self-energy consisting of the scalar and vector parts as
\be
\Sigma_i=\Sigma_i^S-\gamma^\mu\Sigma_{i\mu}^V\, .
\ee
The formal solution of the Schwinger-Dyson equation is
\be
[G_i^H(p)]^{-1} = \gamma\cdot\bar{p}-m_i^\ast
\ee
or equivalently,
\bea
\label{fullbprop}
G_i^H(p) &=& \left(\gamma^\mu\bar{p}_\mu+m_i^\ast\right)
             \Bigg\{\frac{1}{\bar{p}^2 -{m_i^\ast}^2+i\epsilon}\no \\
&&+ \frac{\pi i}{E_i^\ast(p)}\left[\frac{\delta(\bar{p}^0
    -E_i^\ast(p))}{{\rm e}^{\beta(E_i^\ast(p)-\mu_i^\ast)}+1}
    +\frac{\delta(\bar{p}^0+E_i^\ast(p))}{
    {\rm e}^{\beta(E_i^\ast(p)+\mu_i^\ast)}+1}\right]\Bigg\}\no \\
&\equiv & (G_i^H)^F(p) + (G_i^H)^D(p)\,
\eea
where $E_i^\ast(p)=\sqrt{{\bf p}^2+{m_i^\ast}^2}$, $\bar{p}=p+\Sigma_i^V$
and $m_i^\ast=m_i+\Sigma_i^S$.\\
In the present investigation for the study of hot hyperonic matter,
the baryons couple to both the non-strange ($\sigma$) and strange
($\zeta$) scalar fields, so that we have the scalar self energy as
\be
\Sigma^S_i=-(g_{\sigma i}\tilde{\sigma}+g_{\zeta
i}\tilde{\zeta})\, ,
\ee
where $\tilde{\sigma}=\sigma-\sigma_0$, $\tilde{\zeta}=\zeta-\zeta_0$.
The scalar self-energy $\Sigma^S_i$ can also be written as
\bea
\label{scalarsen}
\Sigma^S_i &=& i\left(\fr{g_{\sigma i}^2}{m_\sigma^2}
 + \fr{g_{\zeta i}^2}{m_\zeta^2}\right)\int\fr{\rm{d}^4p}{(2\pi)^4}
 \Tr\big[G_i^F(p)+G_i^D(p)\big]e^{ip^0\eta}\no \\
&\equiv& \big(\Sigma^S_i\big)^F + \big(\Sigma^S_i\big)^D \, .
\eea
$ (\Sigma^S_i)^D $ is the density dependent part and is identical
to the mean field contribution
\bea
\label{denpart}
\big(\Sigma^S_i\big)^D & = & -\left(\fr{g_{\sigma i}^2}{m_\sigma^2}
+ \fr{g_{\zeta i}^2}{m_\zeta^2}\right)\times \gamma_i
\int\fr{\rm{d}^3p}{(2\pi)^3}
\fr{m_i^\ast}{E_i^\ast(p)}\, (f_i(p)+\bar f_i(p)) \no \\
&=& -\left(\fr{g_{\sigma i}^2}{m_\sigma^2}+\fr{g_{\zeta i}^2}
    {m_\zeta^2}\right)\rho_i^S \, .
\eea
The Feynman part $ (\Sigma^S_i)^F $ of the scalar part of the self-energy
is divergent. We adopt a dimensional regularization scheme to extract
the convergent part by performing the integration in n dimensions.
Finally one takes the limit $n \rightarrow 4$ to extract
the divergence which is rendered finite by adding the appropriate
counter terms. After regularization, we finally get
\be
\label{regscalselfen}
(\Sigma^S_i)^F= \fr{\gamma_i}{8\pi^2}\left(\fr{g_{\sigma i}^2}{m_\sigma^2}
+\fr{g_{\zeta i}^2}{m_\zeta^2}\right)\left[{m_i^\ast}^3\Gamma(1-n/2)
+2{m_i^\ast}^3\ln{m_i^\ast}\right],
\ee
with $m_i^\ast = m_i-g_{\sigma i}\tilde{\sigma}-g_{\zeta
i}\tilde{\zeta}$.
We renormalize the Feynman part of the self energy, as given by
the first term on the rhs. of (\ref{scalarsen}) by adding the counter term
\be
\label{ctc}
\left(\Sigma^S_i\right)_{CTC} = - \left(\fr{g_{\sigma i}^2}{m_\sigma^2}
+\fr{g_{\zeta i}^2}{m_\zeta^2}\right)\sum_{n=0}^3\fr{1}{n!}
(g_{\sigma i}\tilde{\sigma}+g_{\zeta i}\tilde{\zeta})^n\beta_{n+1}^{i}\,.
\label {selfctc}
\ee
The coefficients $\beta_n^{i}$'s are fixed from a set of 
renormalization conditions. The first term in (\ref{selfctc})
ensures that the tadpole contribution vanishes in vacuum.
The second term fixes the masses for the scalar fields
$\sigma$ and $\zeta$ fields to their vacuum values, in addition to
ensuring that there are no mixed terms (of the type $\sigma \zeta$)
introduced due to such RHA contributions. The last two terms
in (\ref{selfctc}) are chosen so that there are no cubic or
quartic interaction contributions in the scalar meson fields
arising from contributions in RHA in vacuum.
Explicitly, the coefficients $\beta_n^{i}$'s are given as
\bea
\beta_1^{i} &=& \fr{\gamma_i}{8\pi^2}
\big[m_i^3\Gamma(1-n/2)+2m_i^3\ln{m_i}\big]\, ,
\no \\
\beta_2^{i} &=& - \fr{\gamma_i}{8\pi^2}
\big[3m_i^2\Gamma(1-n/2)+2m_i^2(1+3\ln{m_i})
\big]\, ,
\no \\
\beta_3^{i} &=& \fr{\gamma_i}{8\pi^2}
\big[6m_i\Gamma(1-n/2)+10m_i +12m_i\ln{m_i})\big]\, ,
\no \\
\beta_4^{i} &=& - \fr{\gamma_i}{8\pi^2}
\big[6\Gamma(1-n/2)+22+12\ln{m_i})\big]\, .
\eea
This yields additional contributions
from the Dirac sea to the baryon self energy, as
\bea\lefteqn{
\big(\Sigma^S_i\big)^F{^{finite}} =
\big(\Sigma^S_i\big)^F+\big(\Sigma^S_i\big)_{CTC}
 = \fr{\gamma_i}{4\pi^2}
\left(\fr{g_{\sigma i}^2}{m_\sigma^2}+\fr{g_{\zeta i}^2}{m_\zeta^2}\right) }
\nonumber \\
&\times&\Big[{m_i^\ast}^3\ln\left(\fr{m_i^\ast}{m_i}\right)
+ m_i^2(m_i-m_i^\ast)
- \fr{5}{2}m_i(m_i-m_i^\ast)^2 + \fr{11}{6}(m_i-m_i^\ast)^3\Big]\,.
\eea
The above corresponds to the following counter terms in the Lagrangian
as
\be
{\cal L}_{CTC}=\sum_i\sum_{n=1}^4\frac {\beta_n^{i}}{n!}
(g_{\sigma i}\tilde{\sigma}+g_{\zeta i}\tilde{\zeta})^n,
\ee
where the sum extends over all the baryon species $i$.
The energy density can then be evaluated to
be
\be
\epsilon_{RHA}=\epsilon_{MFT}+\Delta\epsilon\, ,
\ee
with the contribution to the energy density from the Dirac sea as
\bea
\Delta\epsilon & = & -\sum_i \fr{\gamma_i}{16\pi^2}
  \biggl[ {m_i^\ast}^4 \ln \Big( \fr{m_i^\ast}{m_i} \Big)
+ m_i^3 (m_i-m_i^\ast) - \fr{7}{2} m_i^2 (m_i-m_i^\ast)^2 \no \\
&& + \fr{13}{3} m_i (m_i-m_i^\ast)^3 - \fr{25}{12} (m_i-m_i^\ast)^4 \biggr].
\eea
%where $m_i^\ast=-g_{\sigma i}\tilde{\sigma}-g_{\zeta i}\tilde{\zeta}$.\\
The field equations for the scalar meson fields are modified to
\be
\frac{\partial(\Omega/V)}{\partial\Phi}\Bigg|_{RHA} =
\frac{\partial(\Omega/V)}{\partial\Phi}\Bigg|_{MFT}
+\sum_i\frac{\pa m_i^\ast}{\pa\Phi}\Delta \rho^s_i = 0
\quad\mbox{with}\quad \Phi = \sigma, \zeta\,,
\ee
%%%%
where the additional contribution to the nucleon scalar density is
given as
\be
\Delta\rho^s_i = -\fr{\gamma_i}{4\pi^2} \left[ {m_i^\ast}^3\ln\left(
                  \fr{m_i^\ast}{m_i}\right)
+ m_i^2(m_i-m_i^\ast) - \fr{5}{2}m_i(m_i-m_i^\ast)^2 + \fr{11}{6}
  (m_i-m_i^\ast)^3\right].
\ee
These make a refitting of some of the parameters
necessary. First we have to account for the change in the energy and
the pressure, i.e. $g_{N\omega}$ and $\chi_0$ have to be refitted.
Due to a change in $\chi_0$ the parameters $k_0, k_2$ and $k_4$
must be adapted to ensure that the vacuum equations for $\sigma,\zeta$
and $\chi$ have minima at the vacuum expectation values of the fields.
Table \ref{parmfhart} shows the parameters corresponding to the
Mean-field and the Hartree approximations.

%%%% TABELLE %%%%%%%%%%%%%%%%%%%%%%%%%%%%%%%%%%%%%%%%%%%%%%%%%%%%%%%%%%%%%%
%The table shows the parameters corresponding to the Mean-field and the
%relativistic Hartree approximations.
\begin{table}[hc]
\begin{center}
\bt{|c|cc|cc|}
\hline
Parameter & Mean & Field  & Hartree & \\
\hline
$g_4$ & 2.7 & 0 & 2.7 & 0 \\
$k_1$ & 1.4 & 1.4 & 1.4 & 1.4 \\
\hline
$g_{N\omega}$ & 13.24 & 10.74 & 10.61 & 9.45\\
$g_{N\rho}$ & 5.04 & 4.1 & 4.04 & 3.6 \\
$\chi_0 $ & 405.7 & 435.8 & 437.4 & 454.2 \\
$k_3$ & -2.57 & -1.95 & -1.91 & -1.56 \\
\hline
$k_0$ & 2.33 & 2.02 & 2 & 1.86 \\
$k_2$ & -5.55 & -5.55 & -5.55 & -5.55 \\
$k_4$ & -0.23 & -0.23 & -0.23 & -0.24 \\
$m_N^\ast/m_N (\rho_0)$ &0.62 & 0.7 & 0.73 & 0.76\\
$m_\sigma $ & 485 & 578 & 583 & 637 \\
%$K$ & -- & -- & -- & --\\
\hline
\hline
\end{tabular}
\caption{\label{parmfhart} Parameters for the Mean-Field and
the Hartree Fit}
\end{center}
\end{table}
%
%%%%%%%%%%%%%%%%%%%%%%%%%%%%%%%%%%%%%%%%%%%%%%%%%%%%%%%%%%%%%%%%%%%%%%%%

\section{Vector meson masses in the medium}
\label{vmeson}
\subsection{Mass modifications of $\omega$ and $\rho$ -mesons}
\label{omgrhomass}

We now examine how the Dirac sea effects discussed in section
\ref{rha} modify the masses of the vector mesons ($\omega$ and $\rho$)
due to their interaction with the in-medium nucleons.
From (\ref{bvwwlag}), the interaction can be written as
\be
\label{nvwwmftlag}
{\cal L}_{NV}= g_{N\omega} \omega_\mu
 \bar{\psi}_{N} \gamma^\mu \psi_{N} +
 g_{N\rho} \vec{\rho}_\mu
 \bar{\psi}_{N} \gamma^\mu \vec{\tau} \psi_{N},
\ee
in terms of the renormalized couplings $g_{N\omega}$ and $g_{N\rho}$.
Furthermore a tensor coupling is introduced:
\begin{equation}
\label{tensorww}
{\cal L}_{\rm tensor}= - \frac { g_{NV} \kappa_V}{2 m_N}
\left[\bar \psi_N \sigma_{\mu \nu} \tau^a \psi_N \partial ^\nu V^\mu_a \right],
\label{lint}
\end{equation}
\noindent where $(g_{NV},\kappa_V)=(g_{N\omega},\kappa_\omega)$
or $(g_{N\rho},\kappa_\rho)$ for $V_a^\mu = \omega^\mu$ or $\rho_a^\mu$,
$\tau_a =1$ or $\vec{\tau}$, $\vec{\tau}$ being the Pauli matrices.
The vector meson self energy is expressed in
terms of the full nucleon propagator (\ref{fullbprop}) and is given by
\begin{equation}
\Pi_V^{\mu \nu} (k)=-\gamma_I g_{NV}^2 \frac {i}{(2\pi)^4}\int d^4 p\,
{\rm Tr} \Big [ \Gamma_V^\mu (k) G^H(p) \Gamma_V^\nu (-k) G^H(p+k)\Big],
\end{equation}
where $\gamma_I=2$ is the isospin degeneracy factor for nuclear matter,
and, $\Gamma_V^\mu (k)=
\gamma^\mu\tau_a-({\kappa_{V}}/{ 2 m_N})\sigma^{\mu \nu}\tau_a$
represents the meson-nucleon vertex function obtained from (\ref{nvwwmftlag})
and (\ref{tensorww}).
%%%%%%%%%%%%%%%%%%%%%%%%%%%%%%%%%%%%%%%%%%%%%%%%%%%%%%%%%%
For the $\omega$ meson, the tensor coupling is generally small in 
comparison to the vector coupling to the nucleons \cite{hatsuda1} and
is neglected in the present calculations.\\
We then have the meson-nucleon vertex functions as
\bea
\label{nvvertexfn}
\Gamma^\mu(k) &=& \gamma_\mu\, , \quad\quad\quad\quad\quad\quad\mbox{for}
                                 \quad \omega \, ,\no \\
\Gamma^\mu(k) &=& \gamma_\mu+\fr{i\kappa_\rho}{2m_N}\sigma^{\mu\alpha}k_\alpha
\quad\mbox{for}\quad \rho \, .
\eea
After insertion of $G^H(p)$, the vector meson self energy separates into
\be
\Pi^{\mu\nu}(k) = \Pi^{\mu\nu}_F(k) + \Pi^{\mu\nu}_D(k) \, ,
\ee
where $\Pi^{\mu\nu}_F(k)$ is the vacuum polarization and describes the
correction
to the meson propagator due to coupling to baryon-antibaryon excitations
of the Dirac sea and $\Pi^{\mu\nu}_D(k)$ describes coupling to the
particle-hole excitations of the Fermi sea.
The Feynman part of the self energy,  $\Pi^{\mu\nu}_F(k)$, is
divergent and has to be renormalized.
After dimensional regularization to separate the divergent part and
using the subtraction procedure described in \cite{hatsuda,hatsuda1,kurasawa},
we obtain
the following $\Pi^{\mu\nu}_F(k)$ for the $\omega$ and $\rho$ mesons
\bea
\label{vecselfen}
\Pi_F^\omega(k^2) &\equiv& \fr{1}{3}{\rm Re}(\Pi_F^{ren})_\mu^\mu=
-\fr{g_{N\omega}^2}{\pi^2}k^2 I_1 \\
\Pi_F^\rho(k^2) &=& -\fr{g_{N\rho}^2}{\pi^2}k^2
\left[I_1+\fr{m_N^\ast\kappa_\rho}
{2m_N}I_2
+\fr{1}{2}\left(\fr{\kappa_\rho}{2m_N}\right)^2(k^2I_1+m_N^\ast I_2)\right]\, ,
\eea
where
\bea
\label{vecint}
I_1 &=& \int_0^1 dz z (1-z)\ln\Big[\frac {{m_N^*}^2-k^2 z
(1-z)}{{m_N}^2 -k^2 z (1-z)}\Big],
\eea
\bea
I_2 &=& \int_0^1 dz
\ln\Big[\frac{{m_N^*}^2-k^2 z (1-z)}{{m_N}^2 -k^2 z (1-z)}\Big].
\eea
The renormalization condition which has been used to obtain 
the $\omega$-self energy is $\Pi _F^\omega (k^2)(m_N^* \rightarrow 
m_N)=0$ ensuring the vanishing of the vector self energy in vacuum.
Due to the tensor interaction in (\ref{nvvertexfn}), the vacuum
self energy for the $\rho$-meson is not renormalizable. 
We have employed a phenomenological subtraction procedure 
\cite{hatsuda1} to extract the finite part using the condition
$\partial ^n \Pi _F ^\rho (k^2)/\partial (k^2)^n |_{m_N^* \rightarrow
m_N}=0$, with $n=0,1,2,...\infty$.

The contribution from the Fermi sea is given as
\be
\label{vecselfevdp}
\Pi^D (k_0,{\bfm k} \rightarrow 0)
=-\frac {4 g_{NV}^2}{\pi^2}\int p^2 dp\, F(|{\bfm p}|,m_N^*)
\left [  f_{FD}(\mu ^* ,T)+{\bar f}_{FD}(\mu ^* ,T)\right],
\ee
with
\bea
F(|{\bfm p}|,m_N^*) &=& \frac {1}{\epsilon^*(p)(4 \epsilon^*(p)^2-k_0^2)}
\Bigg [ \frac {2}{3} (2 |{\bfm p}|^2+3 {m_N^*}^2)+k_0^2 \Big \{ 2 m_N^*
\Big (\frac {\kappa_V}{2 m_N} \Big) \nonumber \\
&& + \frac {2}{3} \Big ( \frac {\kappa_V}{2 m_N}\Big )^2
(|{\bfm p}|^2+3 {m_N^*}^2)\Big \} \Bigg ],
\eea
where $\epsilon ^* (p)=({\bfm p}^2+{m_N^*}^2)^{1/2}$ is the effective
energy of the nucleon.
The effective mass of the vector meson at rest is then obtained by solving
the equation, with $\Pi =\Pi_F +\Pi_D$,
\be
k_0^2-m_V^2 + {\rm Re} \Pi (k_0,{\bfm k}=0) =0.
\ee
\subsection{Mass modification of $\phi$ - meson}
\label{vphimass}

The $\phi$-meson, which does not couple to the nucleons,
is modified due to the hyperon-antihyperon excitations in the
relativistic Hartree approximation.
The mass of the $\phi$-meson in the medium is obtained as a solution
of the dispersion relation given as \cite{vecmass},
\begin{equation}
k_0^2-m_V^2 +\sum _i {\rm Re} \Pi_i (k_0, \vec k=0) =0.
\label {phimass}
\end{equation}

Using the hyperon-$\phi$ couplings as given in (\ref{gvb})
the mass of the $\phi$-meson is calculated.

\section{Results and discussions}
\begin{figure}
%\begin{center}
\parbox[b]{8cm}{
\includegraphics[width=9.2cm,height=9cm]{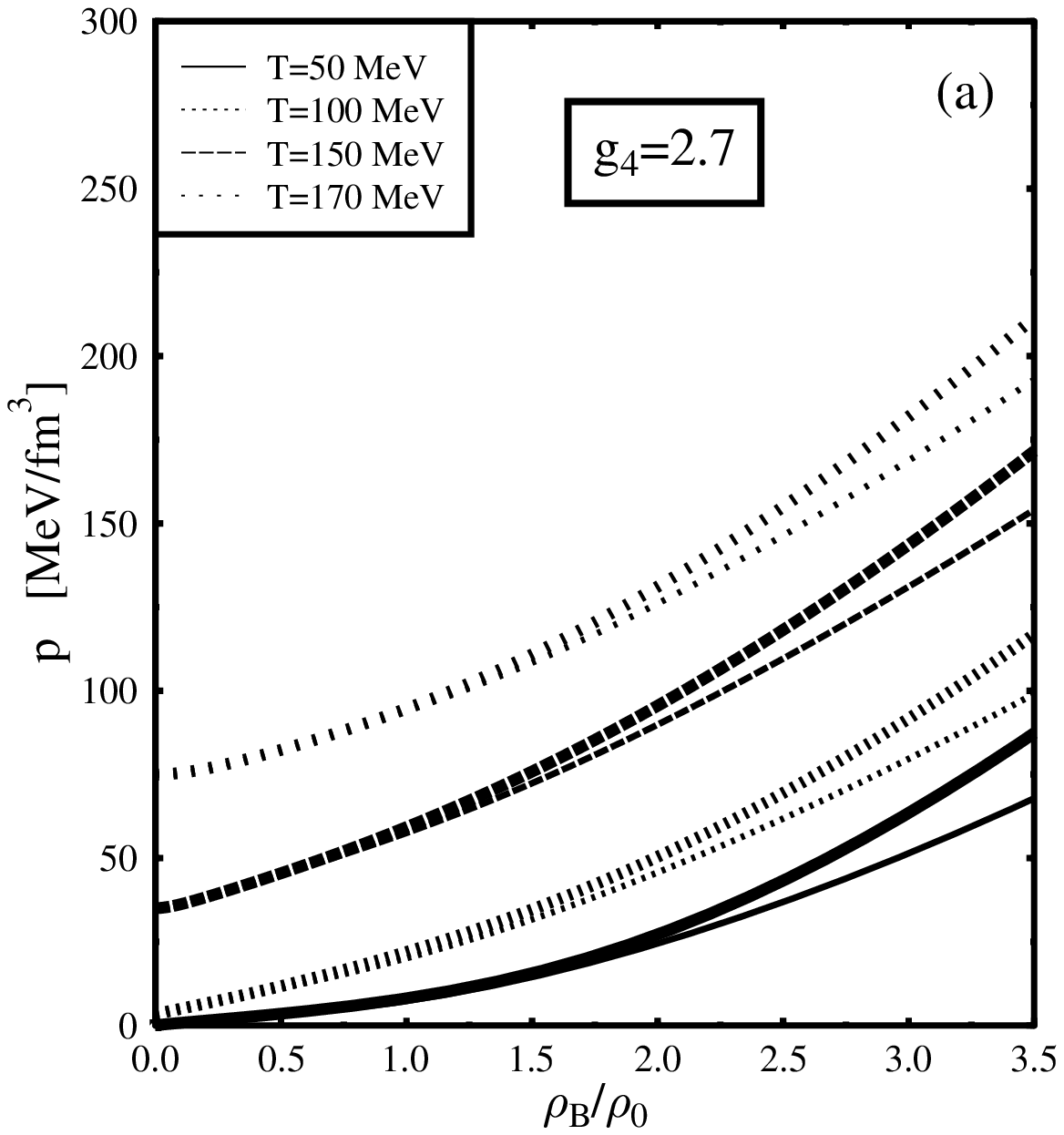}}
\parbox[b]{8cm}{
\includegraphics[width=9.2cm,height=9cm]{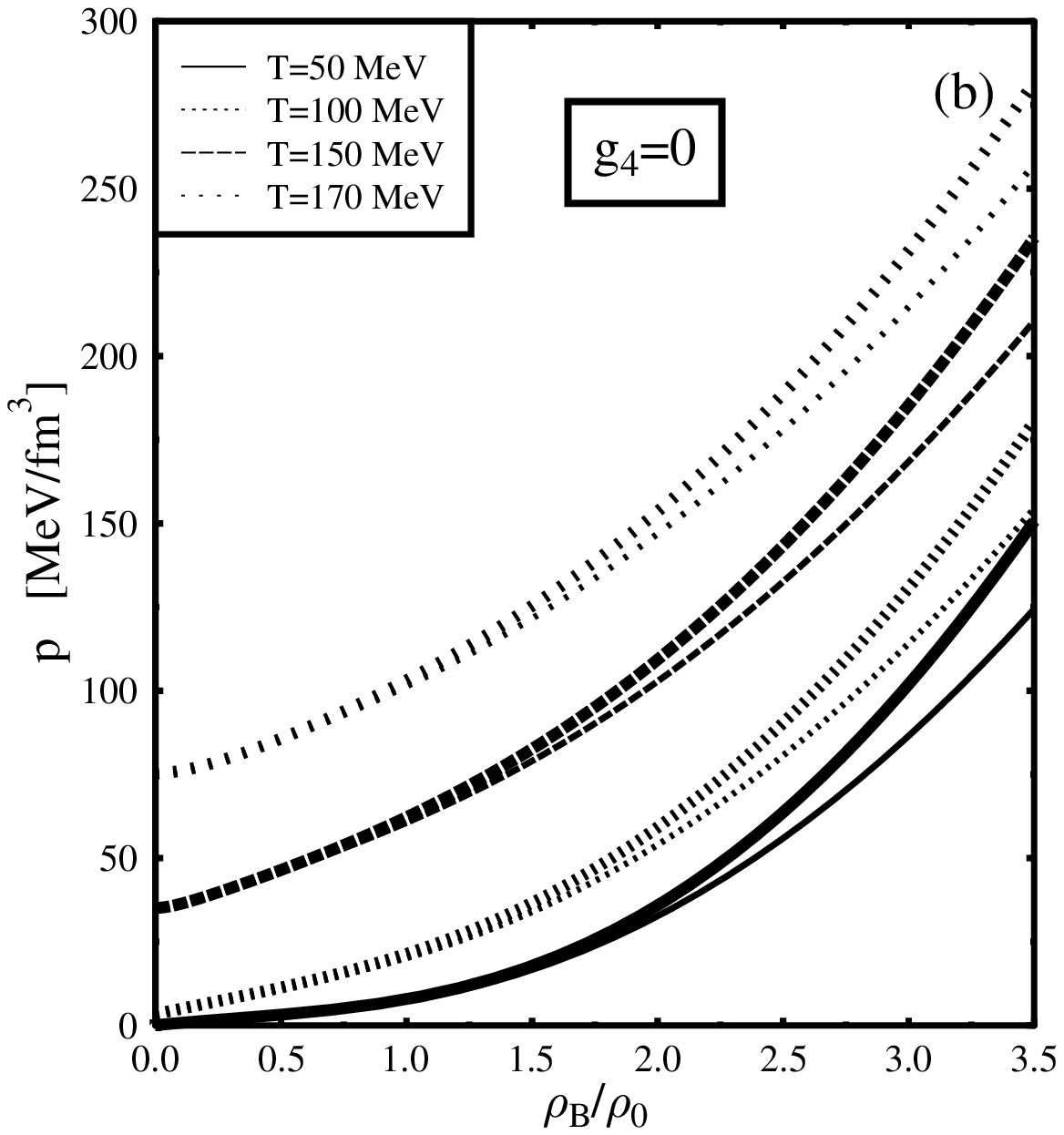}}
\caption{
\label{pre}
The equation of state in the mean field (MFT) (thick lines) and in RHA
(thin lines) for different temperatures and $ f_s = 0 $ for the vector 
self-interaction coupling (a) $g_4$=2.7 and (b) $g_4$=0.}
%\end{center}
\end{figure}
\begin{figure}
%\begin{center}
\parbox[b]{8cm}{
\includegraphics[width=9.2cm,height=9cm]{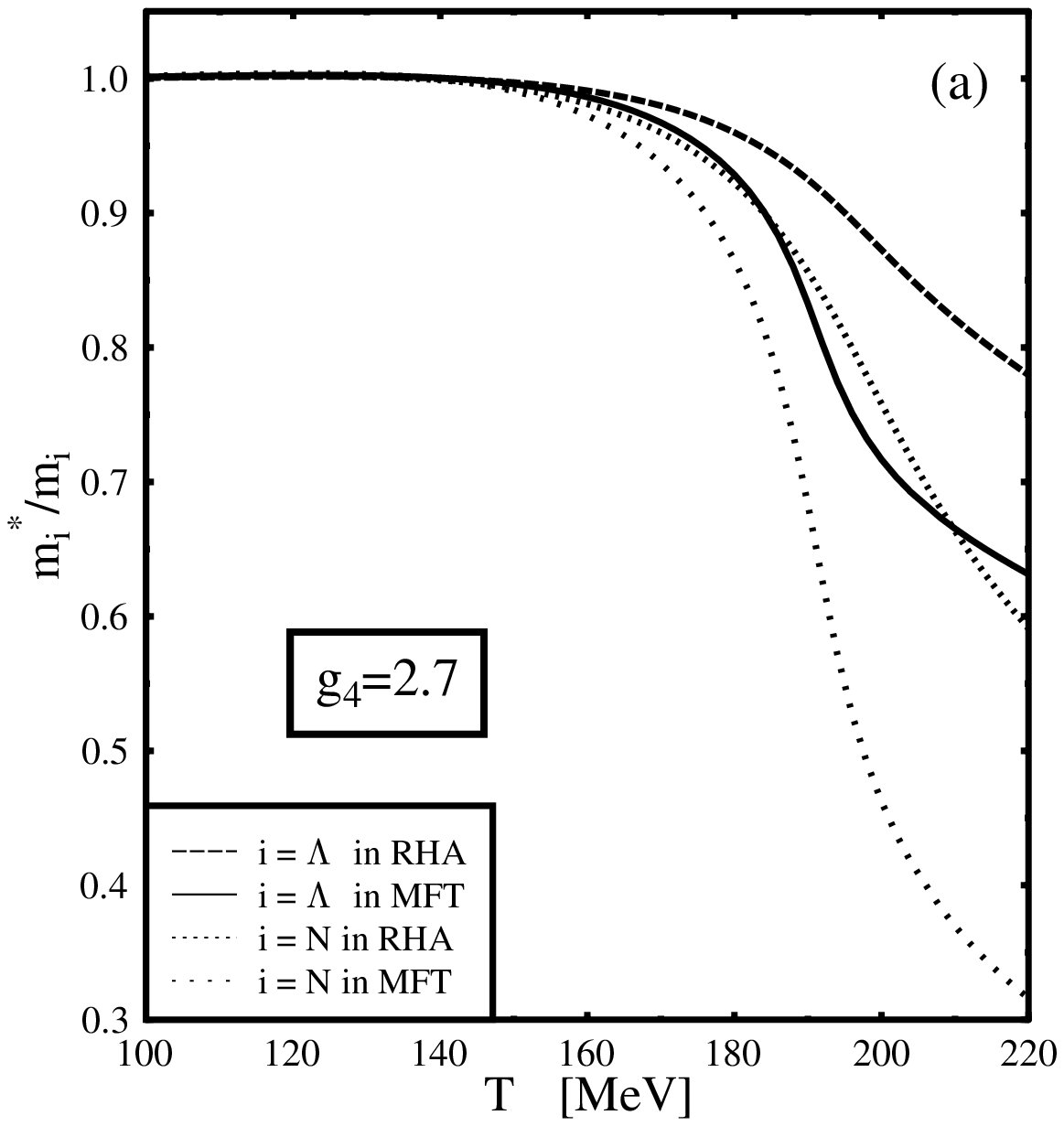}}
\parbox[b]{8cm}{
\includegraphics[width=9.2cm,height=9cm]{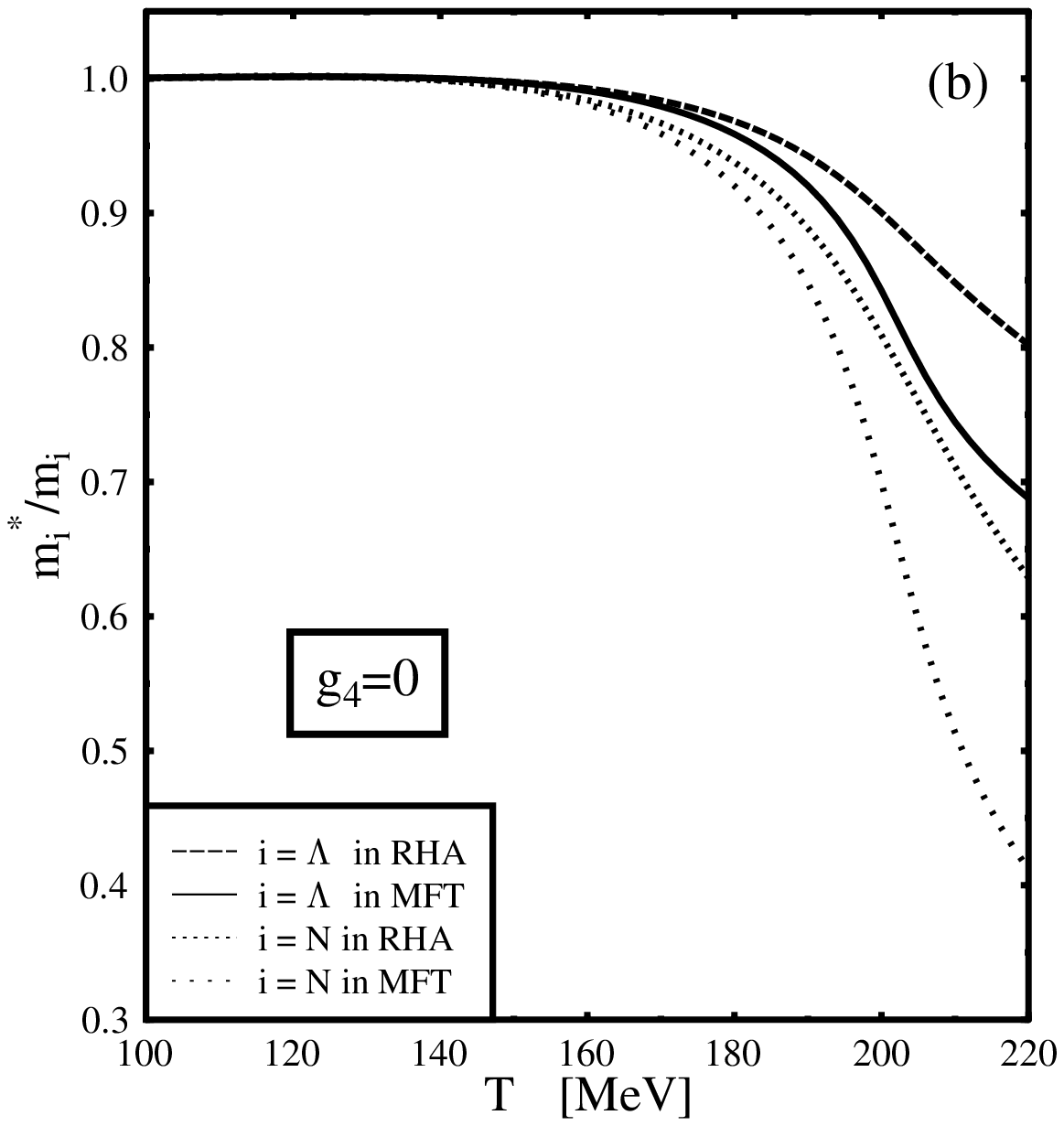}}
\caption{
\label{mnrh0}
Effective nucleon and $\Lambda$ masses as functions of temperature in the
mean field (MFT) and in RHA 
for $ f_s = 0 $ and $ \rho_B = 0 $
and for (a) $g_4$=2.7 and (b) $g_4$=0.}
%\end{center}
\end{figure}
\begin{figure}
%\begin{center}
\parbox[b]{8cm}{
\includegraphics[width=9.2cm,height=9cm]{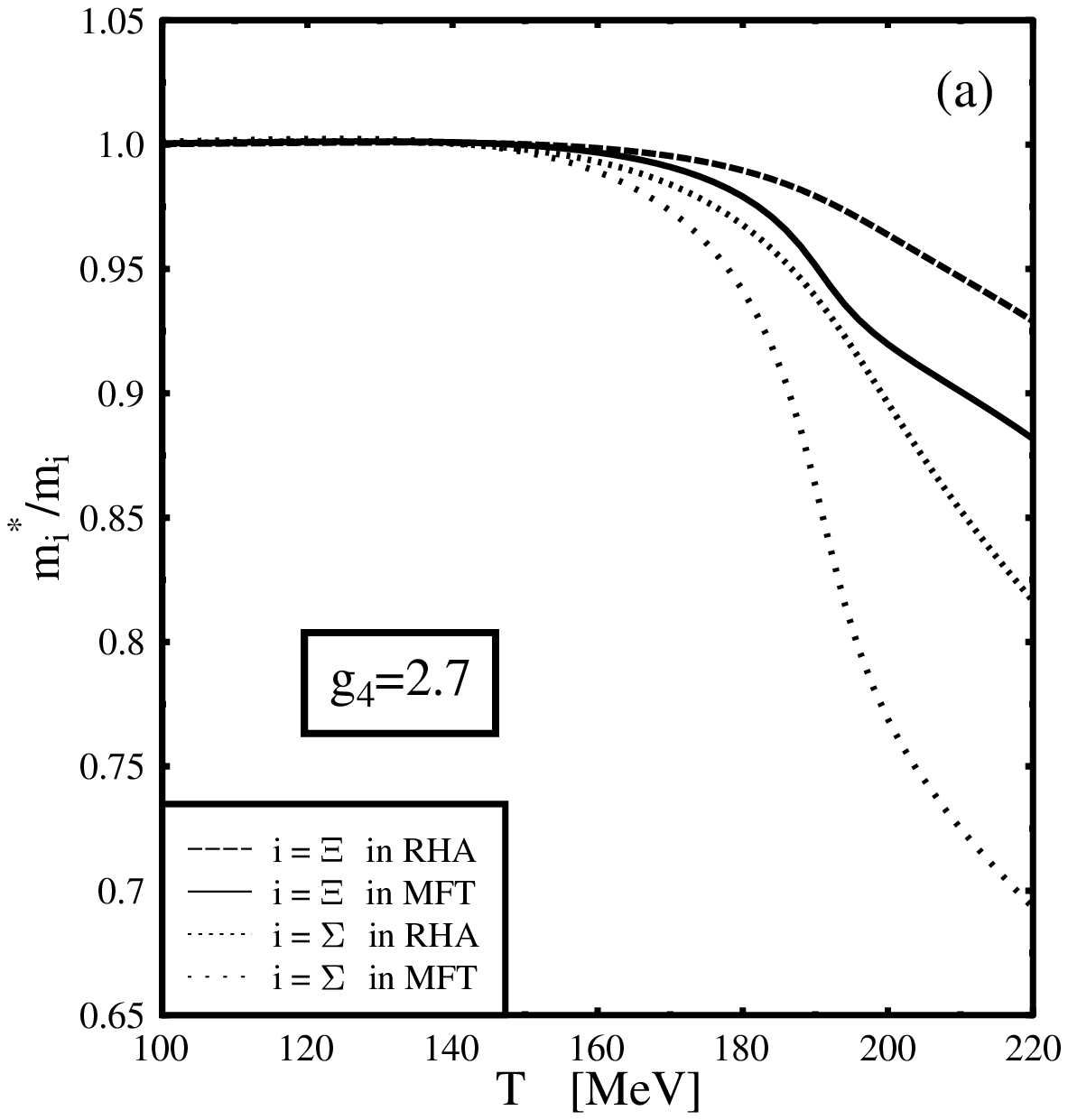}}
\parbox[b]{8cm}{
\includegraphics[width=9.2cm,height=9cm]{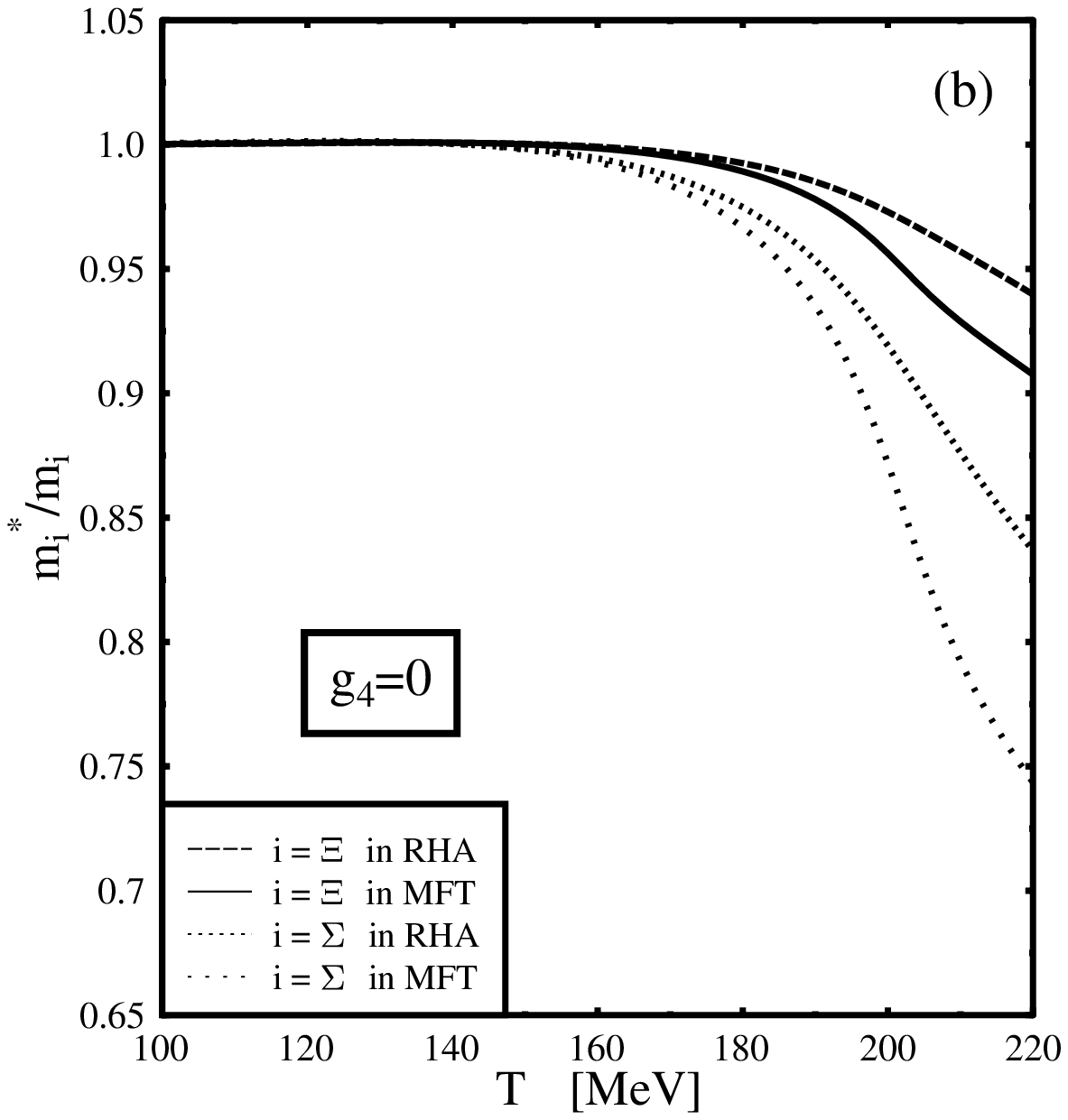}}
\caption{
\label{mhrh0}
Effective $\Sigma$ and $\Xi$ masses as functions of temperature in the
mean field (MFT) and in RHA 
for $ f_s = 0 $ and $ \rho_B = 0 $
and for (a) $g_4$=2.7 and (b) $g_4$=0.}
%\end{center}
\end{figure}
\begin{figure}
%\begin{center}
\parbox[b]{8cm}{
\includegraphics[width=9.2cm,height=9cm]{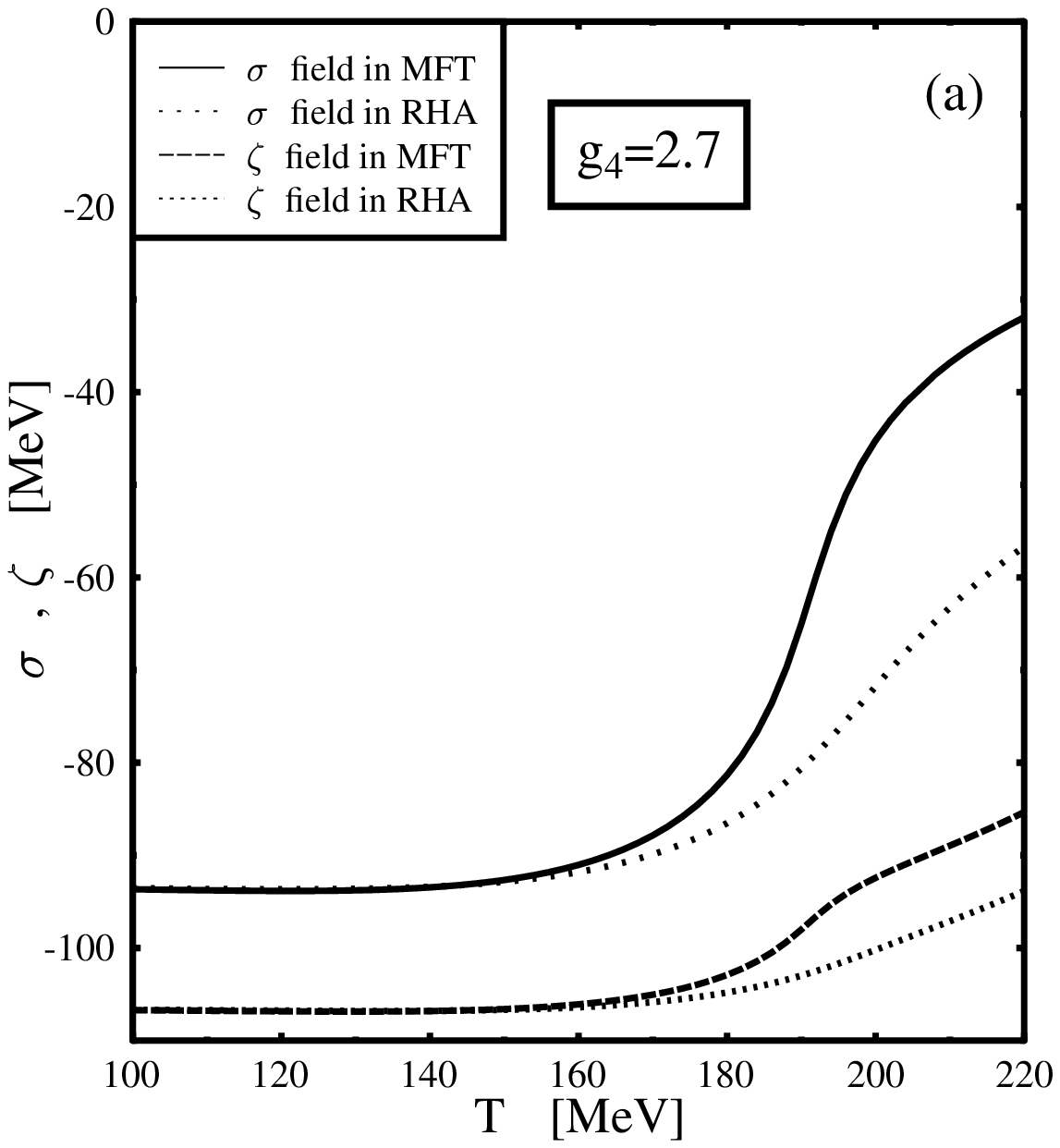}}
\parbox[b]{8cm}{
\includegraphics[width=9.2cm,height=9cm]{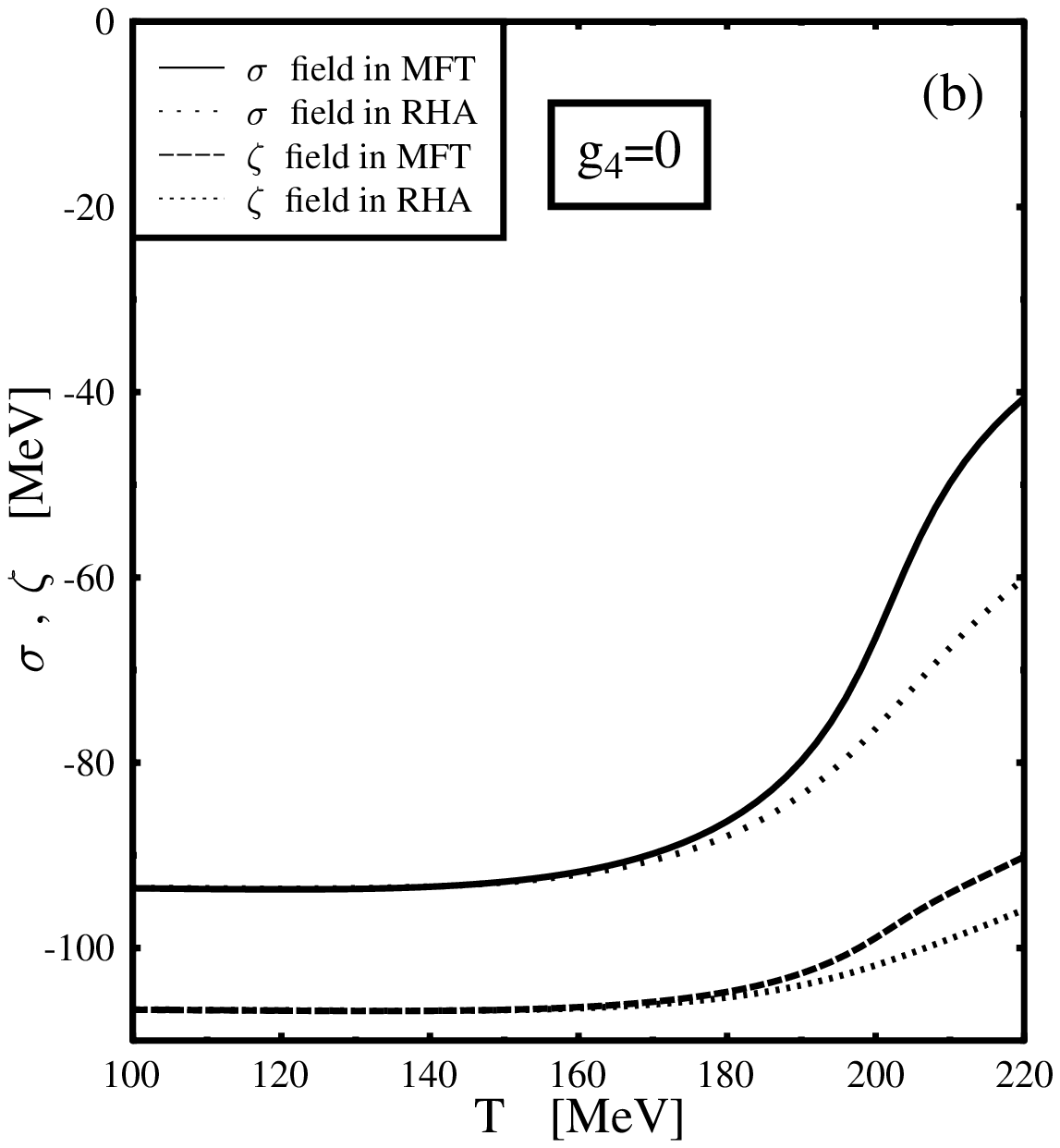}}
\caption{
\label{smesonrh0}
The scalar fields as functions of temperature in the
mean field (MFT) and in RHA 
for $ f_s = 0 $ and $ \rho_B = 0 $
and for (a) $g_4$=2.7 and (b) $g_4$=0.}
%\end{center}
\end{figure}
\begin{figure}
%\begin{center}
\parbox[b]{8cm}{
\includegraphics[width=9.2cm,height=9cm]{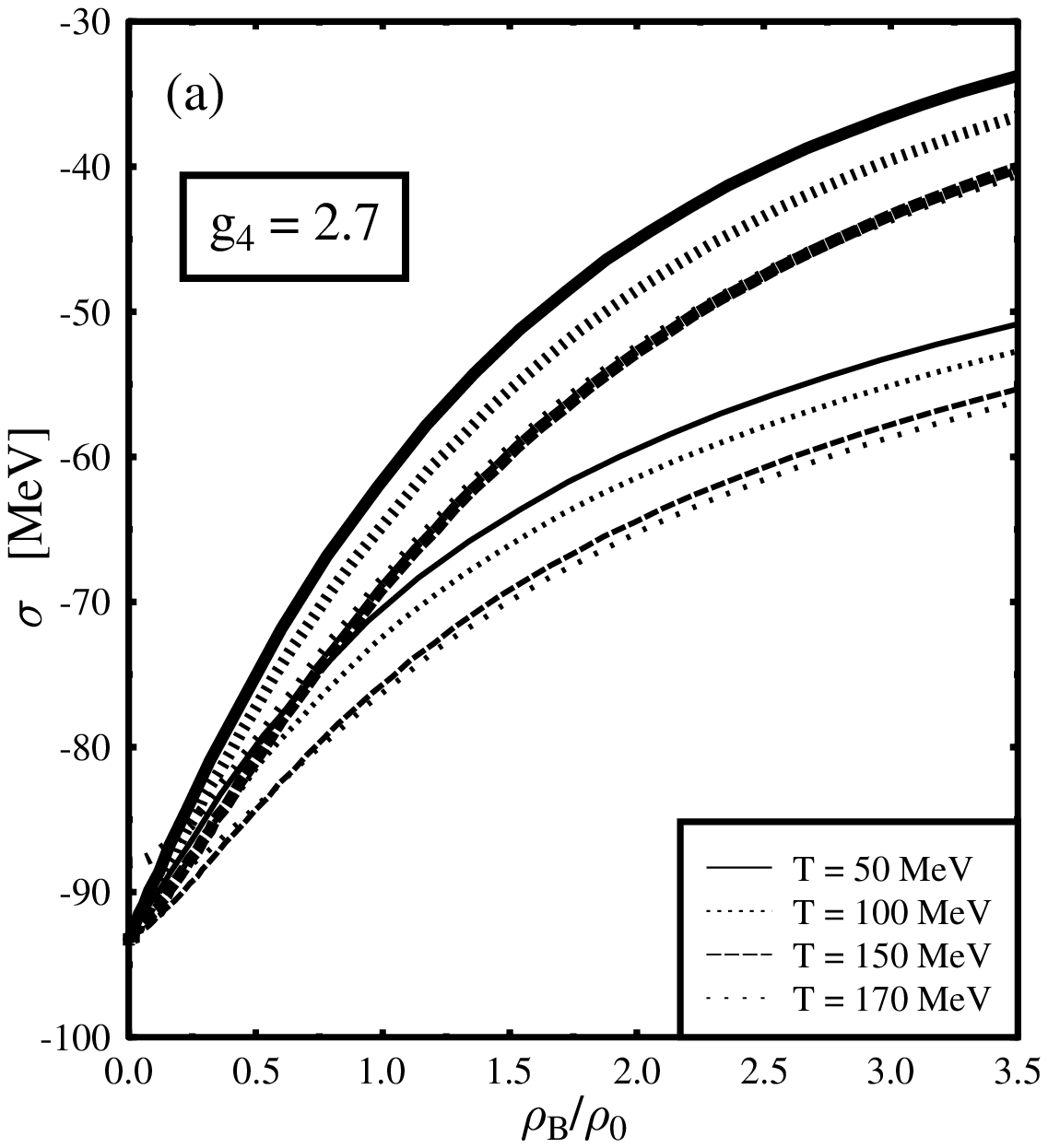}}
\parbox[b]{8cm}{
\includegraphics[width=9.2cm,height=9cm]{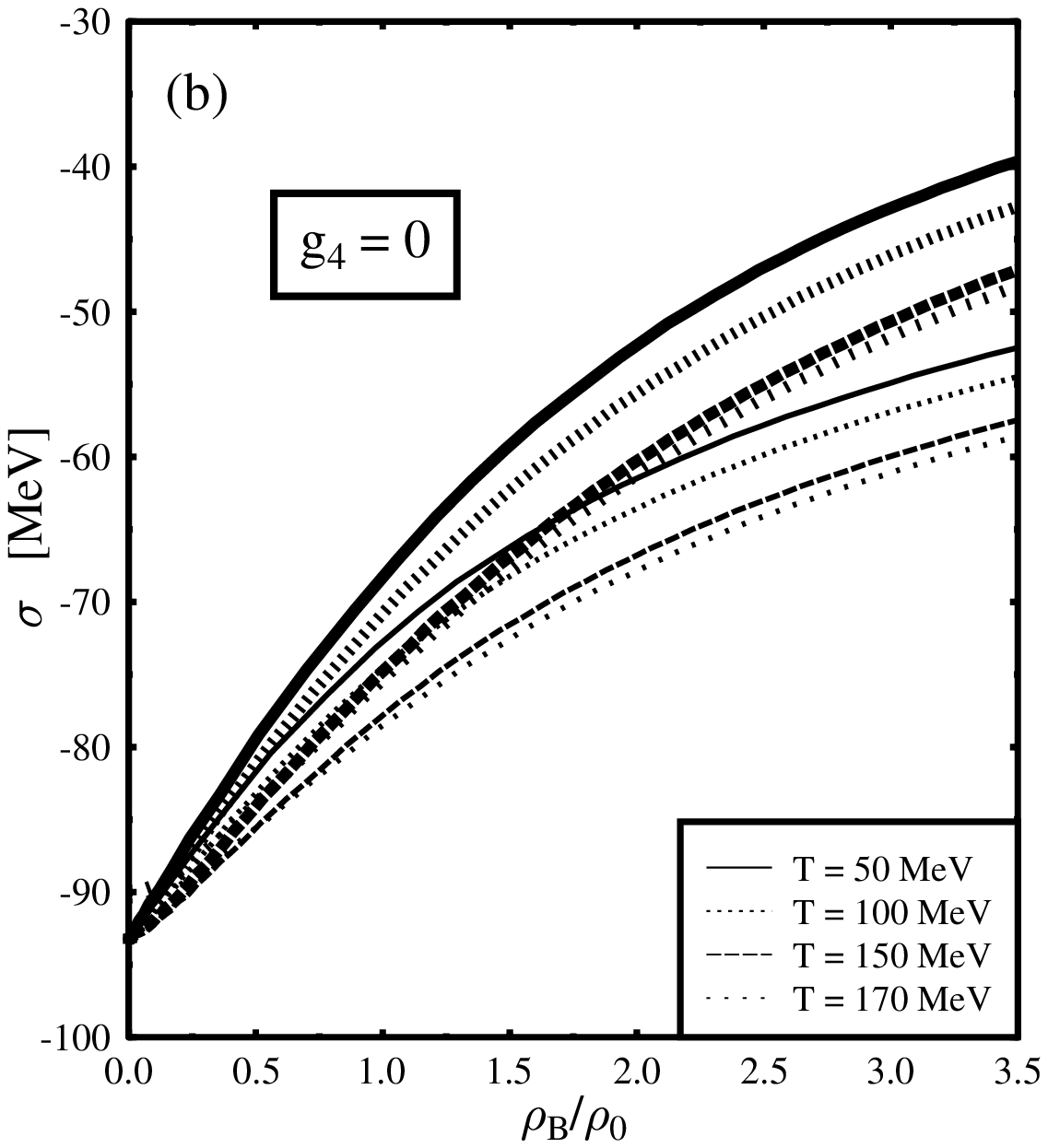}}
\caption{
\label{sgrhtemp}
The non-strange condensate as a function of density in the mean field
(MFT) (thick lines) and in RHA (thin lines)
for different temperatures and $ f_s = 0 $
and for (a) $g_4$=2.7 and (b) $g_4$=0.}
%\end{center}
\end{figure}
\begin{figure}
%\begin{center}
\parbox[b]{8cm}{
\includegraphics[width=9.2cm,height=9cm]{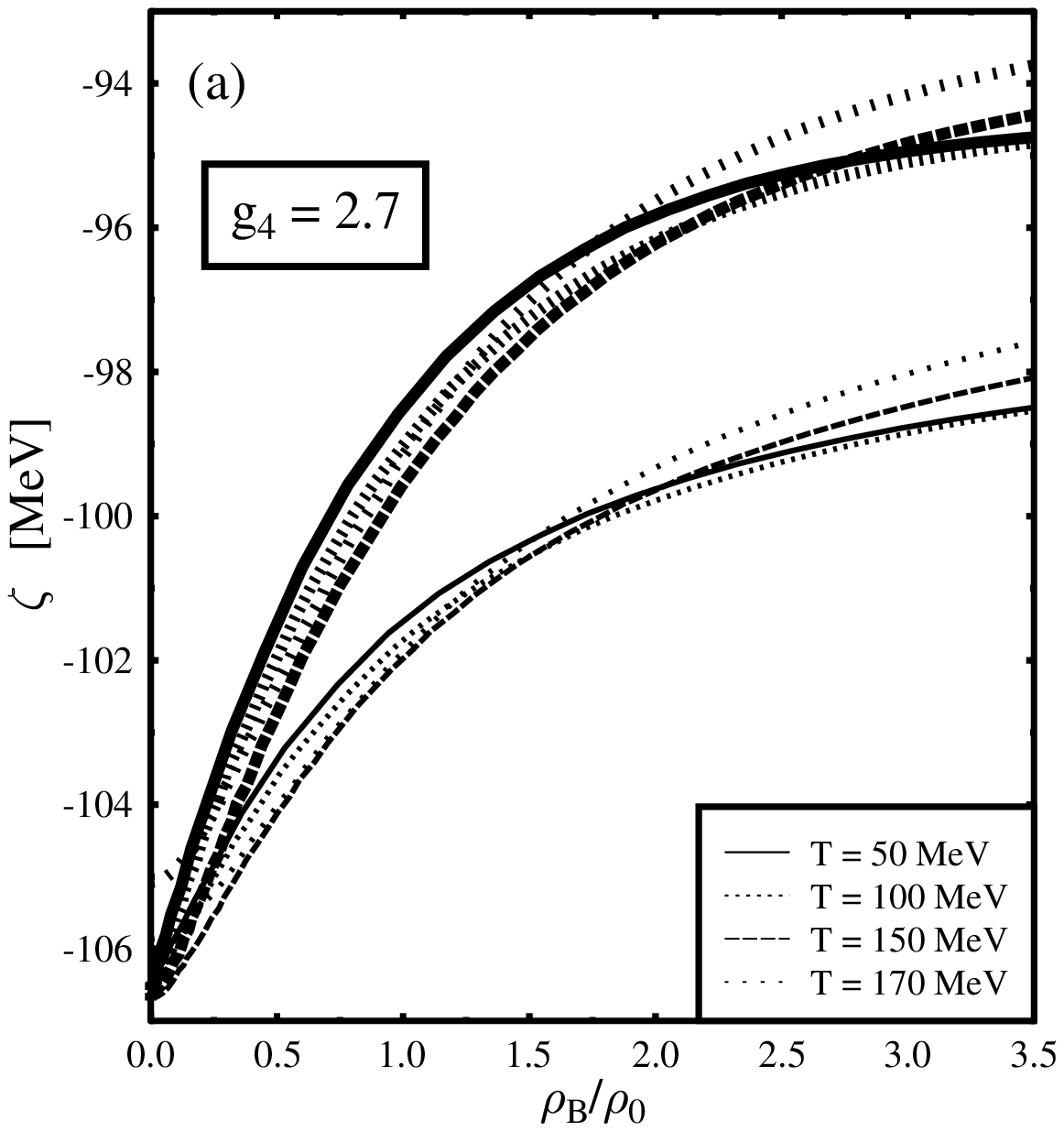}}
\parbox[b]{8cm}{
\includegraphics[width=9.2cm,height=9cm]{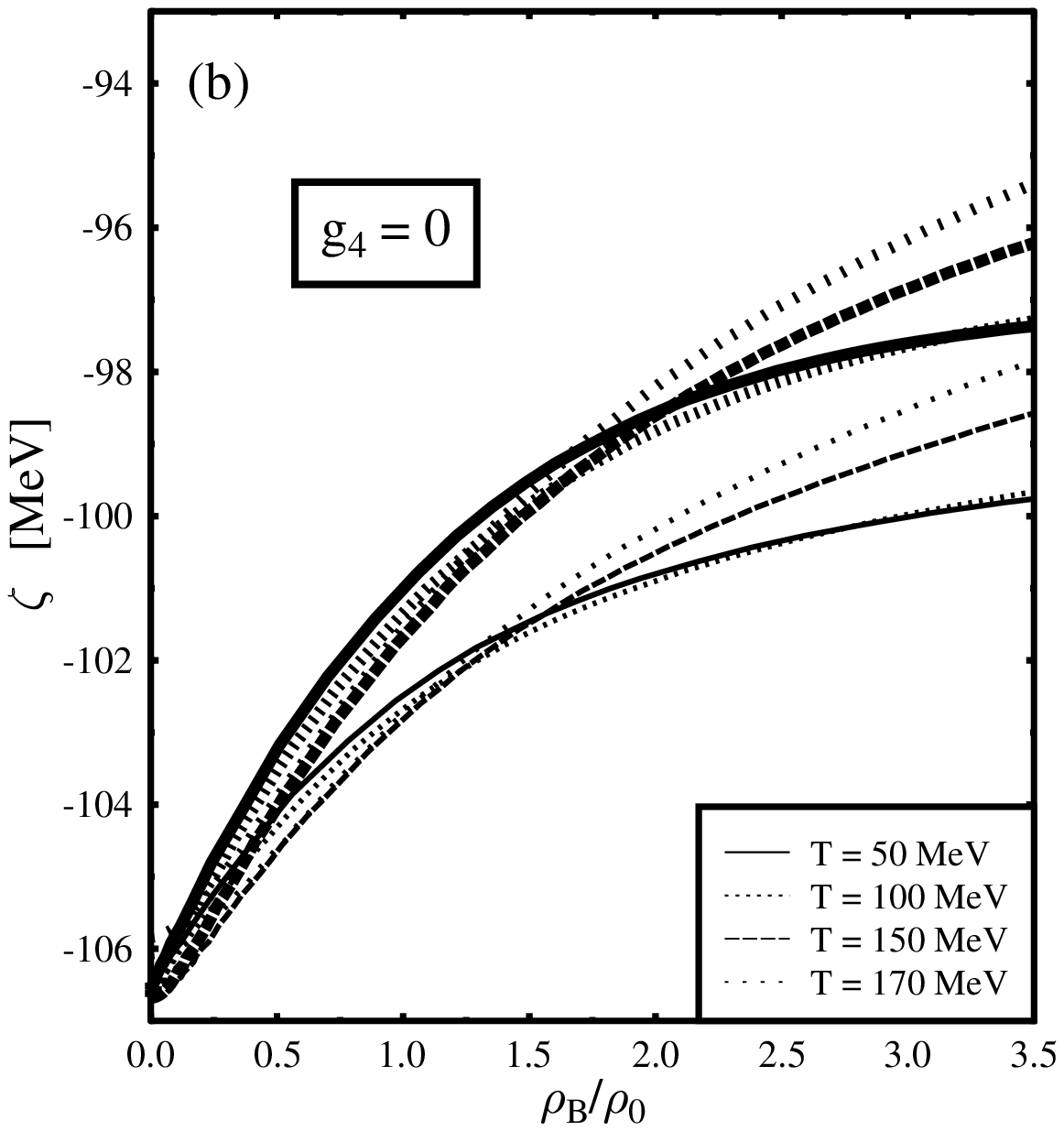}}
\caption{
\label{zetarhtemp}
The strange condensate as a function of density in the mean field
(MFT) (thick lines) and in RHA (thin lines)
for different temperatures and $ f_s = 0 $
and for (a) $g_4$=2.7 and (b) $g_4$=0.}
%\end{center}
\end{figure}
\begin{figure}
%\begin{center}
\parbox[b]{8cm}{
\includegraphics[width=9.2cm,height=9cm]{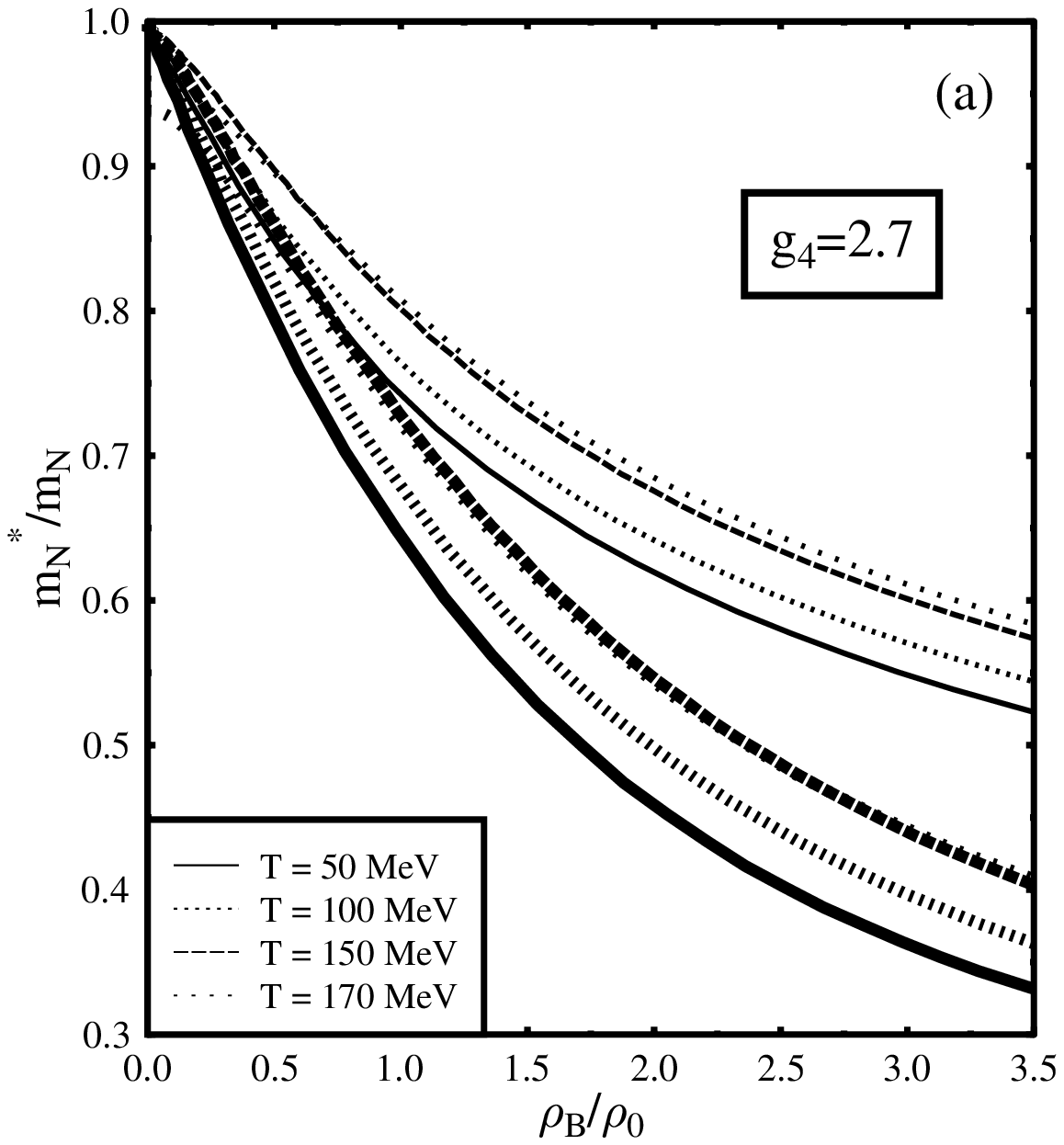}}
\parbox[b]{8cm}{
\includegraphics[width=9.2cm,height=9cm]{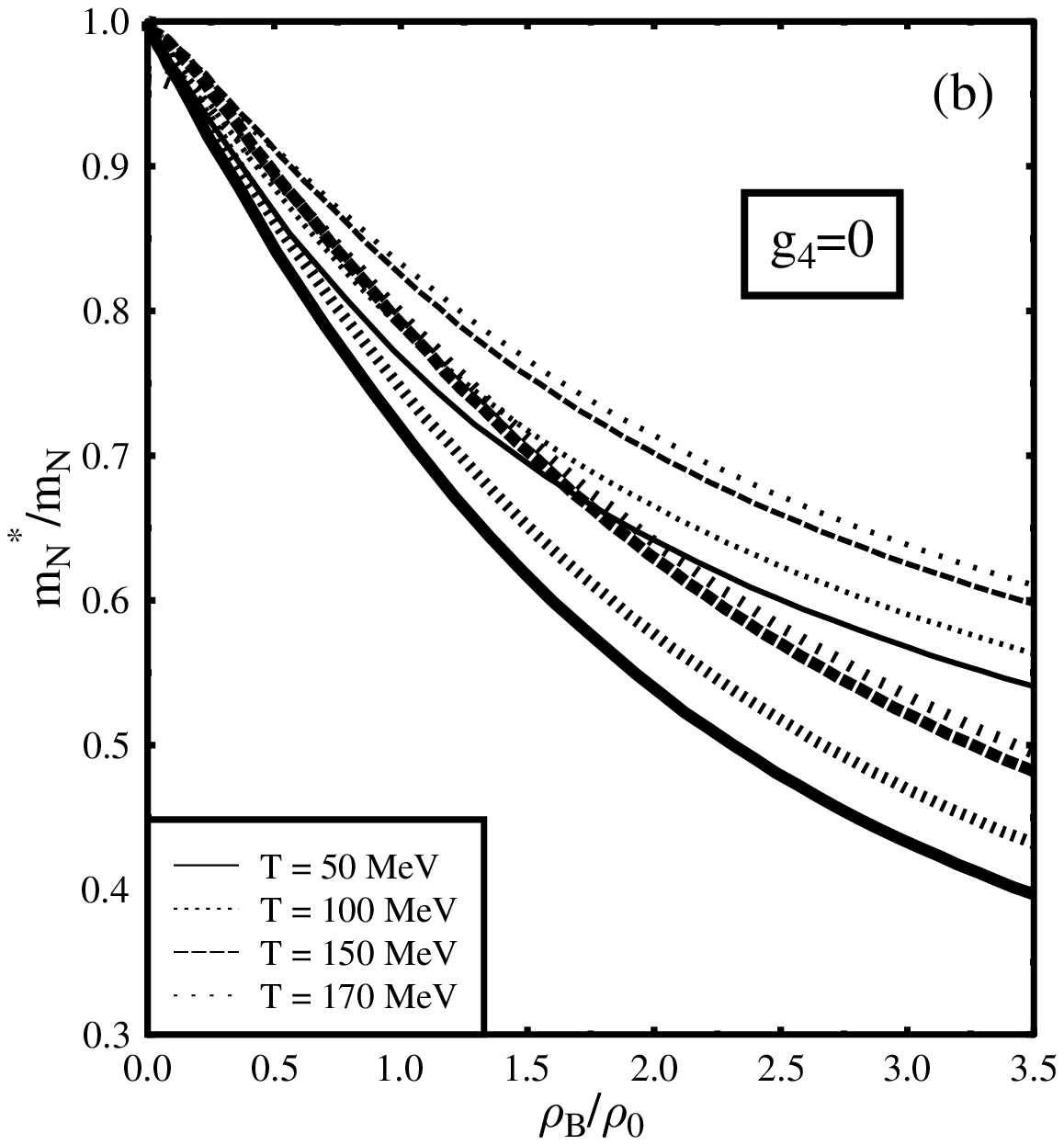}}
\caption{
\label{mnrhtemp}
Effective nucleon mass as a function of density in the mean field
(MFT) (thick lines) and in RHA (thin lines)
for different temperatures and $ f_s = 0 $
for (a) $g_4$=2.7 and (b) $g_4$=0.}
%\end{center}
\end{figure}
\begin{figure}
%\begin{center}
\parbox[b]{8cm}{
\includegraphics[width=9.2cm,height=9cm]{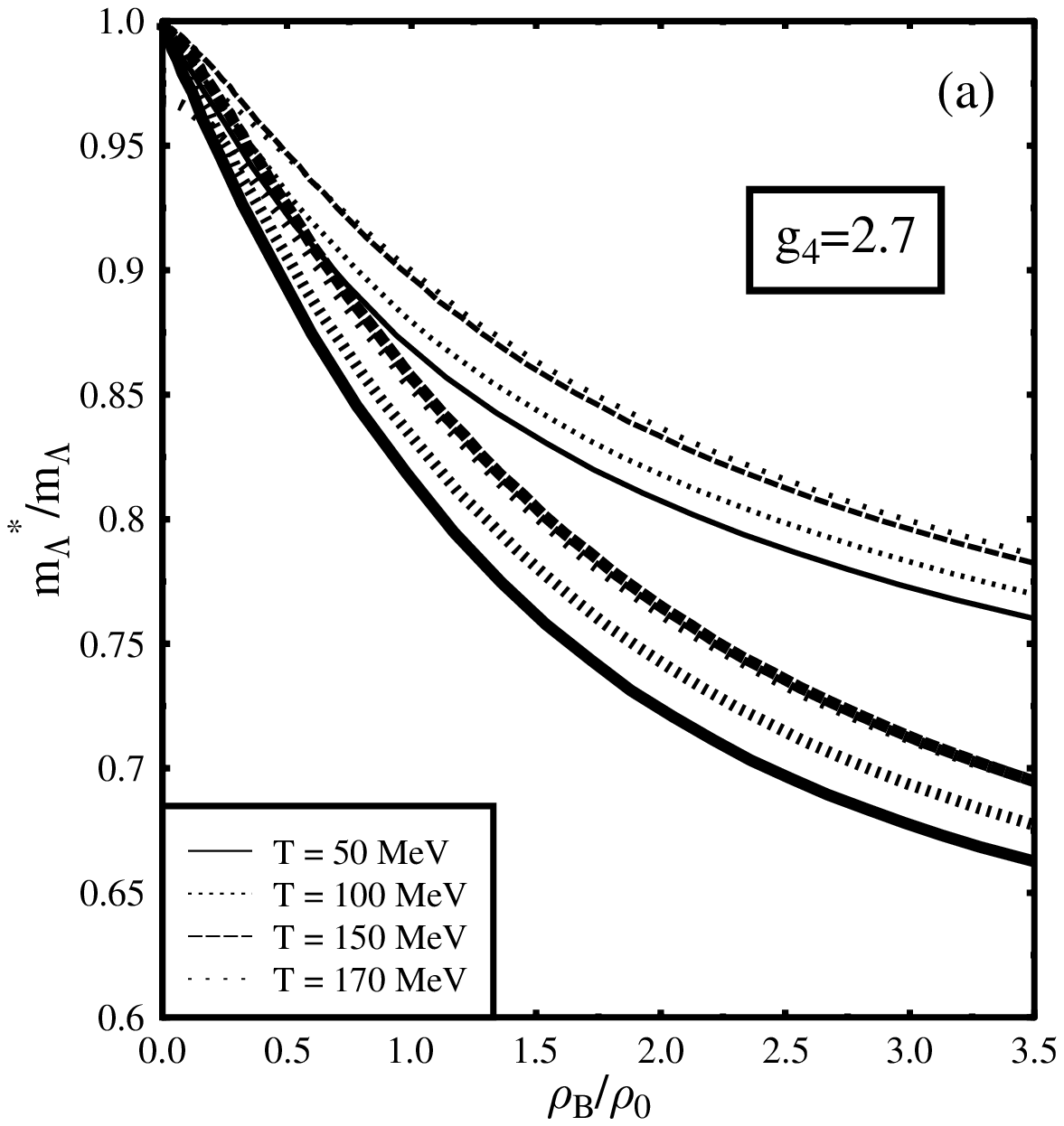}}
\parbox[b]{8cm}{
\includegraphics[width=9.2cm,height=9cm]{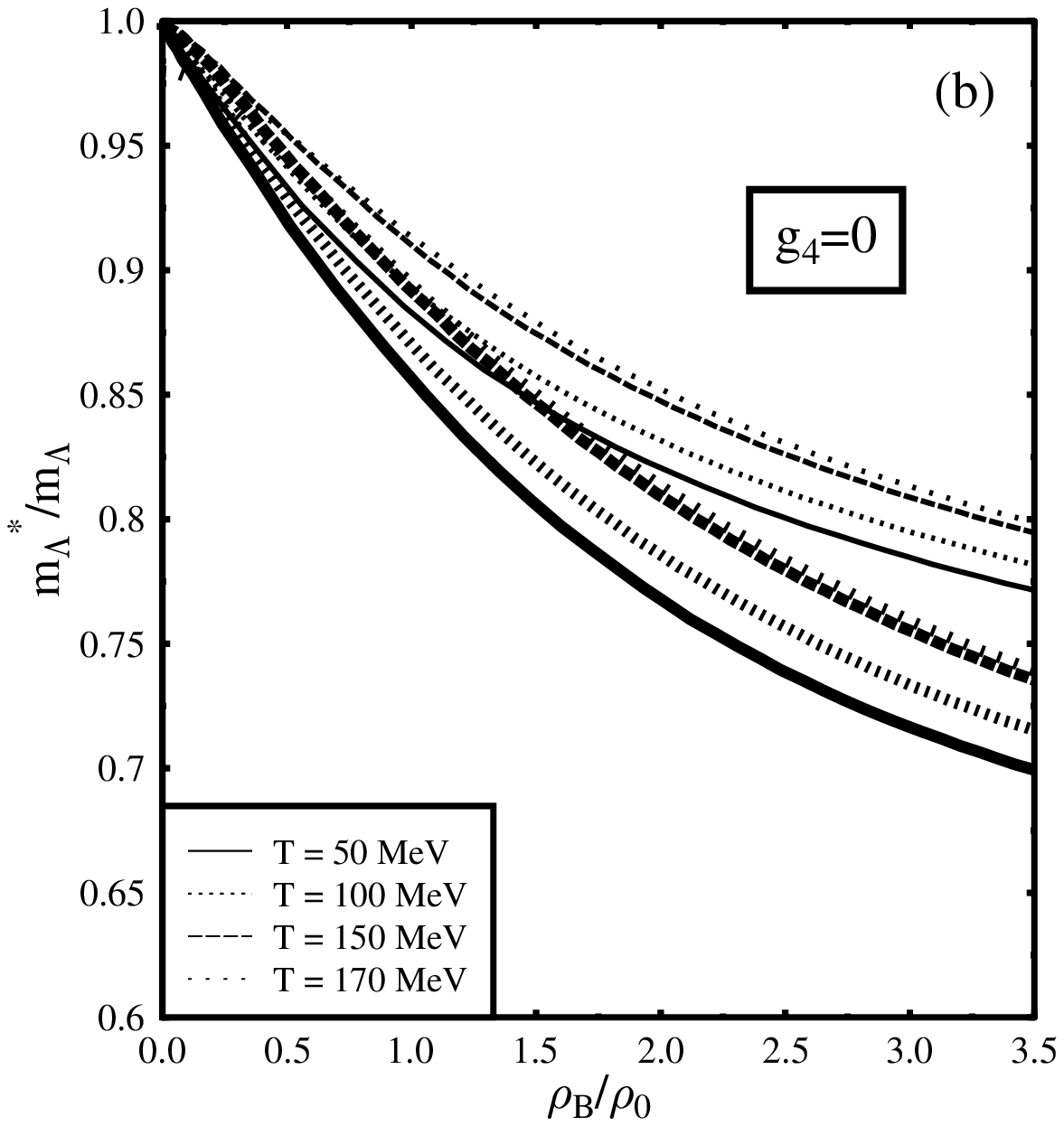}}
\caption{
\label{mlrhtemp}
Effective $\Lambda$ mass as a function of density in the mean field
(MFT) and in RHA
for different temperatures and $ f_s = 0 $
for (a) $g_4$=2.7 and (b) $g_4$=0.}
%\end{center}
\end{figure}
\begin{figure}
%\begin{center}
\parbox[b]{8cm}{
\includegraphics[width=9.2cm,height=9cm]{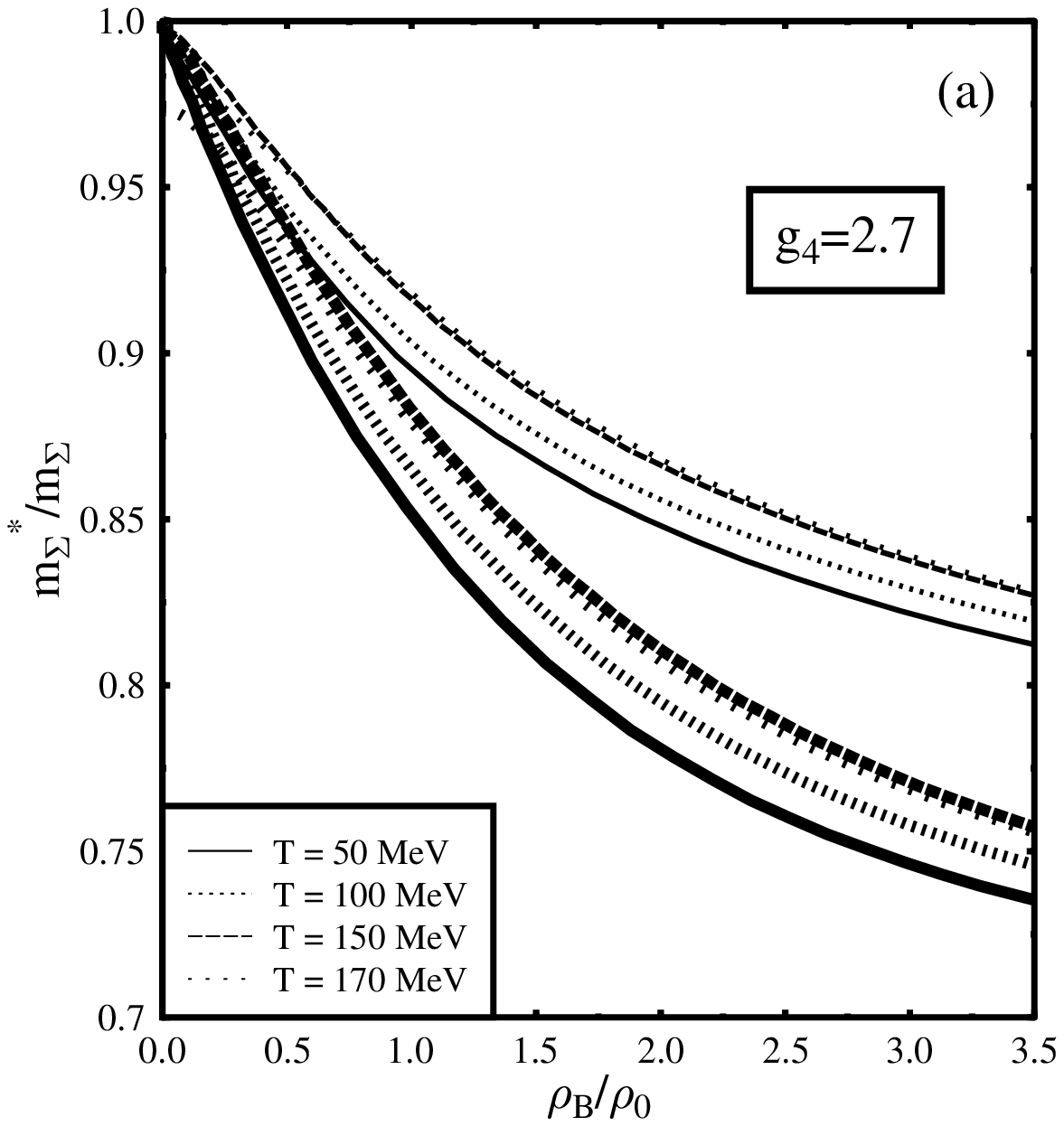}}
\parbox[b]{8cm}{
\includegraphics[width=9.2cm,height=9cm]{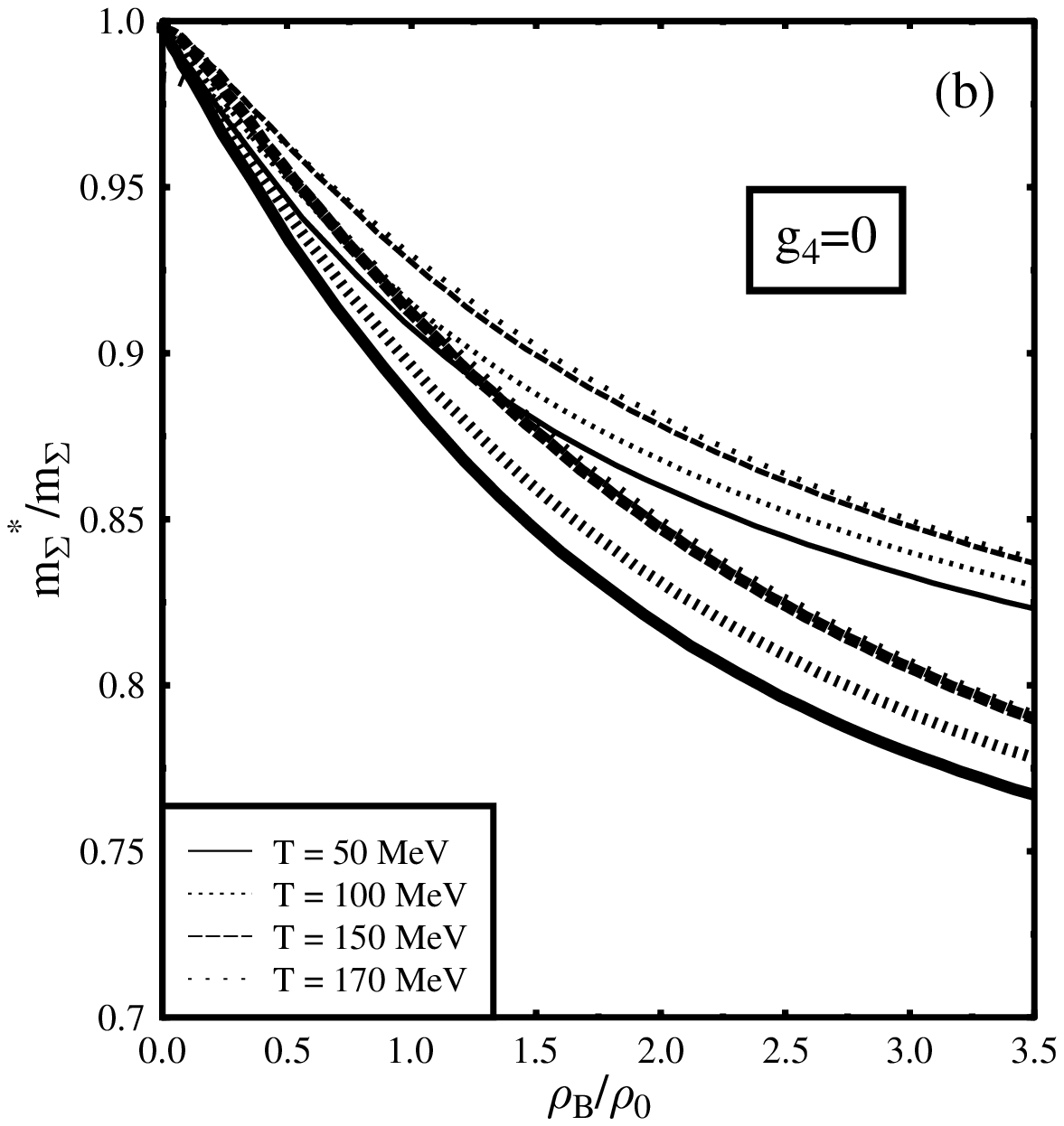}}
\caption{
\label{msgrhtemp}
Effective $\Sigma$ mass as a function of density in the mean field
(MFT) (thick lines) and in 
RHA  (thin lines) for different temperatures and $ f_s = 0 $
for (a) $g_4$=2.7 and (b) $g_4$=0.}
%\end{center}
\end{figure}
\begin{figure}
%\begin{center}
\parbox[b]{8cm}{
\includegraphics[width=9.2cm,height=9cm]{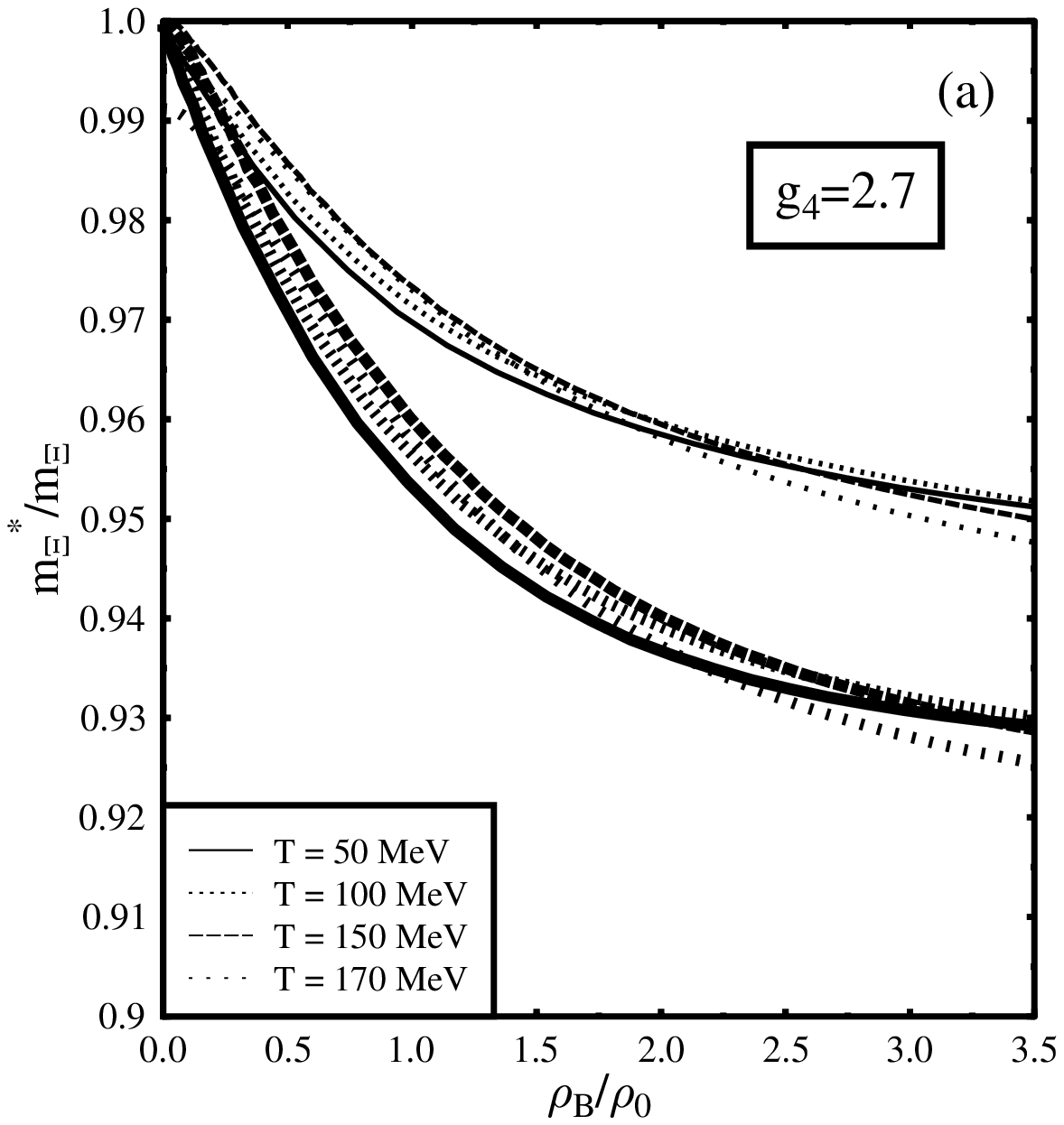}}
\parbox[b]{8cm}{
\includegraphics[width=9.2cm,height=9cm]{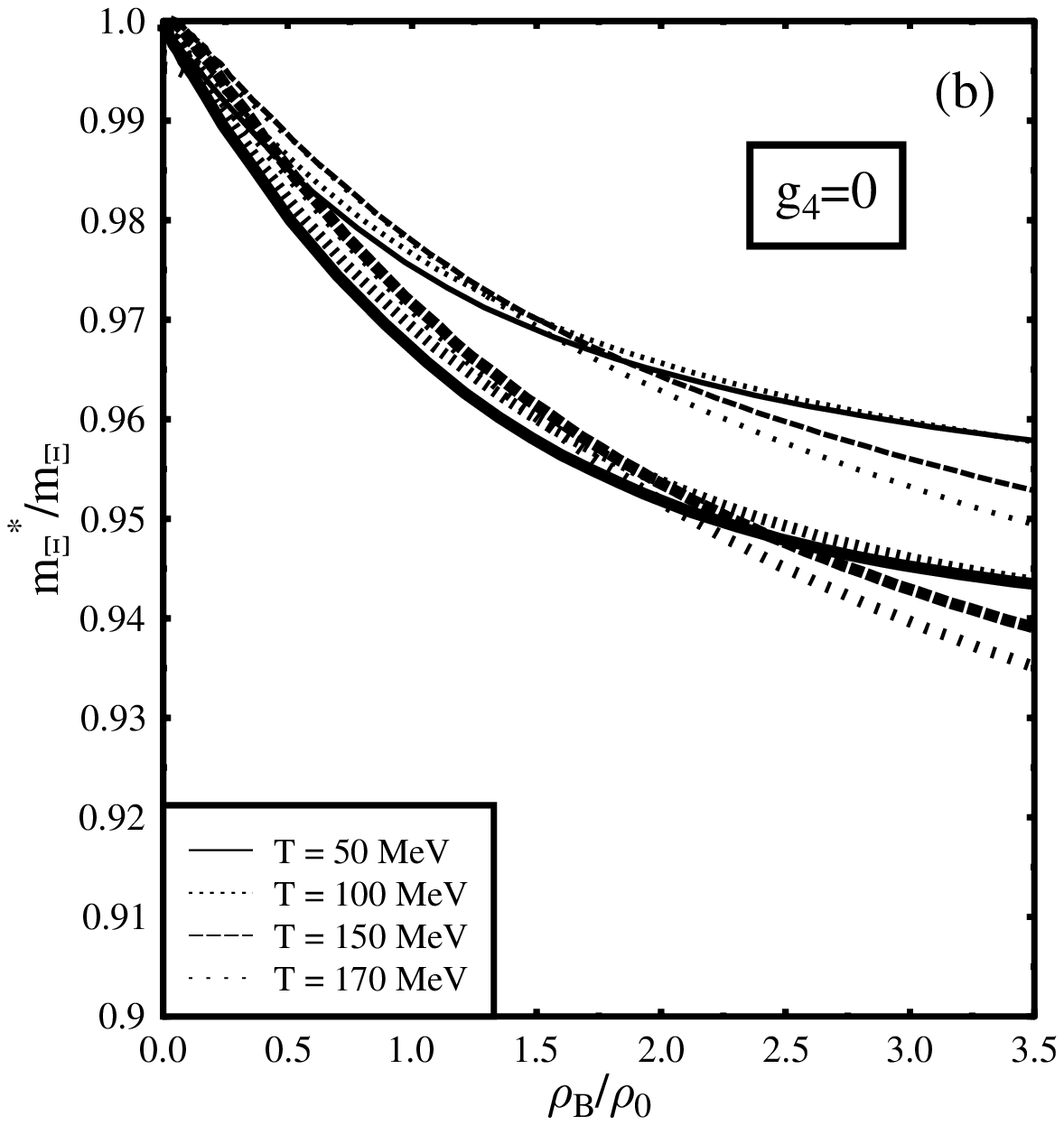}}
\caption{
\label{mxirhtemp}
Effective $\Xi$ masses as a function of density in the mean field
(MFT) (thick lines) and in RHA (thin lines)
for different temperatures and $ f_s = 0 $
for (a) $g_4$=2.7 and (b) $g_4$=0.}
%\end{center}
\end{figure}

In this section, we discuss the findings of the present investigation
for the properties of the hadrons properties in a chiral SU(3) model. 
We investigate how the hadron properties in mean field approximation 
are modified due to the effect of the vacuum
polarizations in relativistic Hartree approximation.
Figure \ref{pre} shows the temperature and density dependence
of the pressure. The effect from the Dirac sea of the baryons
through the relativistic Hartree approximation is seen to lead
to a softening of the equation of state for the hot hyperonic matter.
In figures \ref{mnrh0} and \ref{mhrh0} the baryon masses are shown
as functions of temperature for zero baryon density. Accounting for the
baryonic Dirac sea effects is seen to give rise to higher values 
for the baryon masses in the medium. This is related to the fact 
that these contributions lead to a softening of
the equation of state as illustrated in the figure \ref{pre}.
However, note that the masses stay almost constant
up to a temperature of around 160 MeV above which there is
a drop of the masses in the hot matter.
The modification of the hyperon masses is smaller as compared to the
nucleon mass because of their stronger coupling to the strange
condensate $\zeta$, which shows much weaker temperature dependence
than the non-strange condensate $\sigma$ as illustrated in
\ref{smesonrh0}. Especially, the $\Xi$ mass is seen to stay almost
constant even up to a temperature of about 180 MeV.
The presence of quartic self interaction for the vector field
enhances the mass modification as seen in figures \ref{mnrh0}
and \ref{mhrh0}. This is because the vector field strength 
is attenuated leading to a larger effective chemical potential
and hence a larger thermal distribution function for the baryon.
As a consequence the contribution to baryon scalar self energy
from the medium dependent part as given by (\ref{denpart}) 
becomes larger. This explains the smaller baryon mass 
for $g_4 \neq 0$ in the mean field approximation. 
The qualitative features remain the same with additional
contributions from $N \bar N$ fluctuations in RHA.
Figures \ref{sgrhtemp} and \ref{zetarhtemp}
illustrate the temperature and density dependence
of the non-strange ($\sigma$) and strange ($\zeta$) scalar fields.
The effective baryon masses as functions of density, for different
temperatures and zero net strangeness are shown in figures
\ref{mnrhtemp} -\ref{mxirhtemp}.
Again, higher baryon masses are predicted if we take the quantum
corrections of the Dirac sea into account.
This behaviour mirrors itself in the density dependence of the
scalar fields (figures \ref{sgrhtemp}-\ref{zetarhtemp}).

Note that the masses of the baryons at finite densities
first increases with temperature up to around 170 MeV and then
decreases. Such a behaviour of the nucleon mass increasing with temperature
was also observed earlier within the framework of the Walecka model
by Ko and Li \cite {liko} in a mean field calculation.
This behaviour of the baryon self energy, given by (\ref{denpart})
in the mean field approximation can be understood in the following manner.
The thermal distribution functions have the effect of increasing
the self energy as given by (\ref{denpart}) (hence decreasing the masses).
However, at finite densities, for increasing temperatures,
there are contributions also from higher momenta thereby
increasing the denominator of the integrand in the rhs of
(\ref{denpart}) (and hence increasing the value for the effective baryon mass).
The competing effects arising from the thermal distribution functions
and the contributions from higher-momenta states give rise to the observed
behaviour of the effective baryon masses with temperature
at finite densities.
For temperatures of about 170 MeV one can see that the
scalar fields have nonzero fluctuations from the vacuum values
even at zero density, which indicates the existence of 
baryon-antibaryon pairs in the thermal bath.  
This behaviour for nuclear matter
 at finite temperatures has also been known in
the literature \cite{furnst,theis}. This leads to the baryon masses
above this temperature to be different from the vacuum values
 at zero baryon density.

%The magnitude of the scalar meson field, $\sigma$ as obtained
%through the self-consistency condition (\ref{selfsig}), is plotted
%as a function of the baryon density for various temperatures
%in figure 6 for $\lambda_R$=1.8.
%It may be noted that for $\rho_B$=0, $\sigma$ field becomes
%nonzero at a temperature of around 160 MeV due to thermal effects,
%and is observed to have attained an appreciable value at 200 MeV
%due to contributions from the thermal distribution functions.
%This nonzero value for the sigma field is associated with
%appearance of thermal nucleon-antinucleon pairs, and, may
%be interpreted as the `boiling' of vacuum.
%The sigma field attaining a nonzero value was also observed
%for nuclear matter in the mean field approximation in
%Walecka model \cite{furnst,theis}, which had led to a sharp fall
%in the effective nucleon mass between 150 MeV and 200 MeV.
%The rapid fall of $M_N^*$ with increasing temperature, was
%ascribed to resemble the situation, when the system becomes
%a dilute gas of baryons in a sea of baryon-antibaryon pairs.

%
\begin{figure}
%\begin{center}
\parbox[b]{8cm}{
\includegraphics[width=9.2cm,height=9cm]{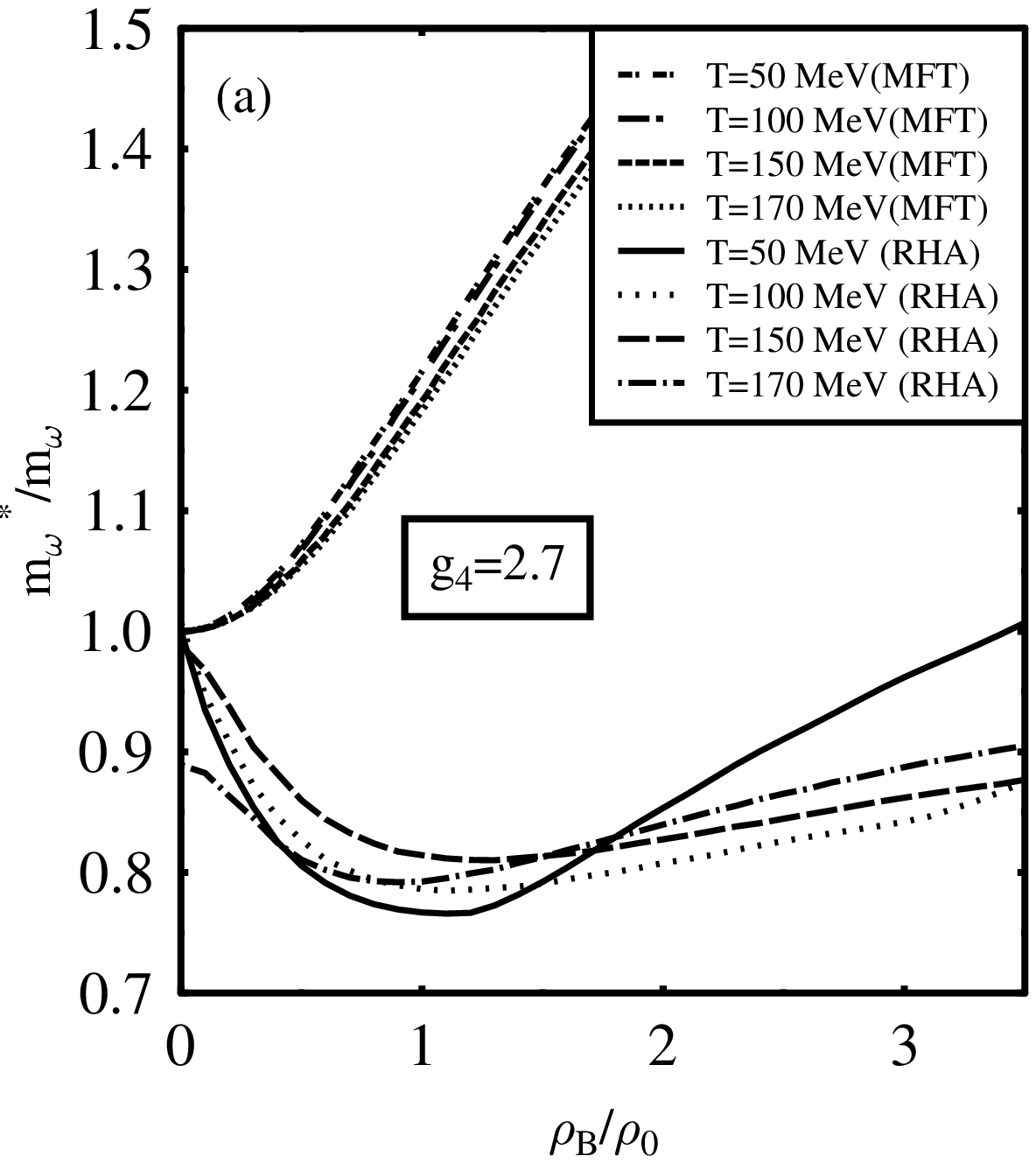}}
\parbox[b]{8cm}{
\includegraphics[width=9.2cm,height=9cm]{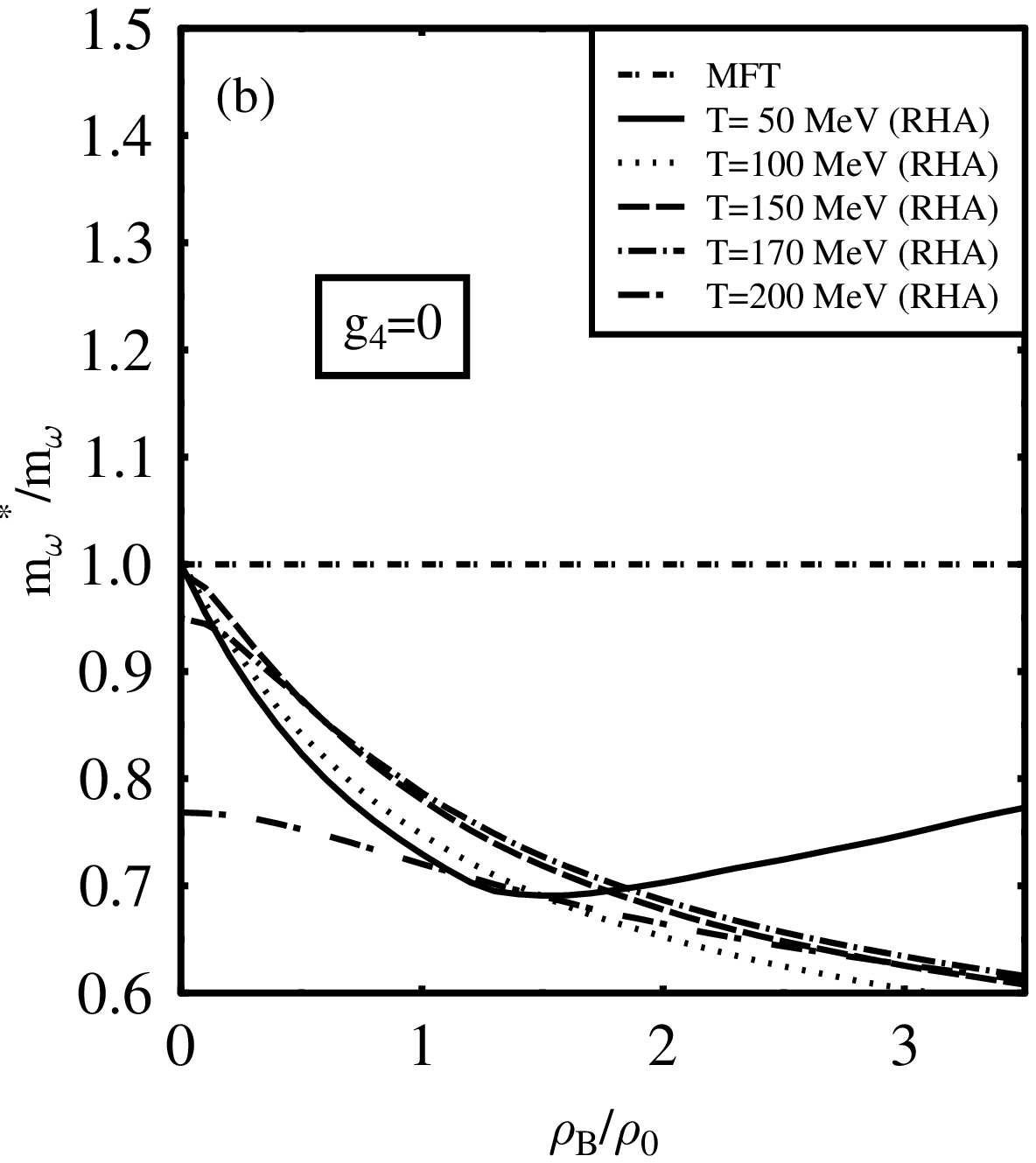}}
\caption{
\label{momgfig}
Effective $\omega$ mass as a function of density in the mean field
(MFT) and in RHA
for different temperatures and $ f_s = 0 $
and for (a) $g_4$=2.7 and (b) $g_4$=0.}
%\end{center}
\end{figure}

\begin{figure}
%\begin{center}
\parbox[b]{8cm}{
\includegraphics[width=9.2cm,height=9cm]{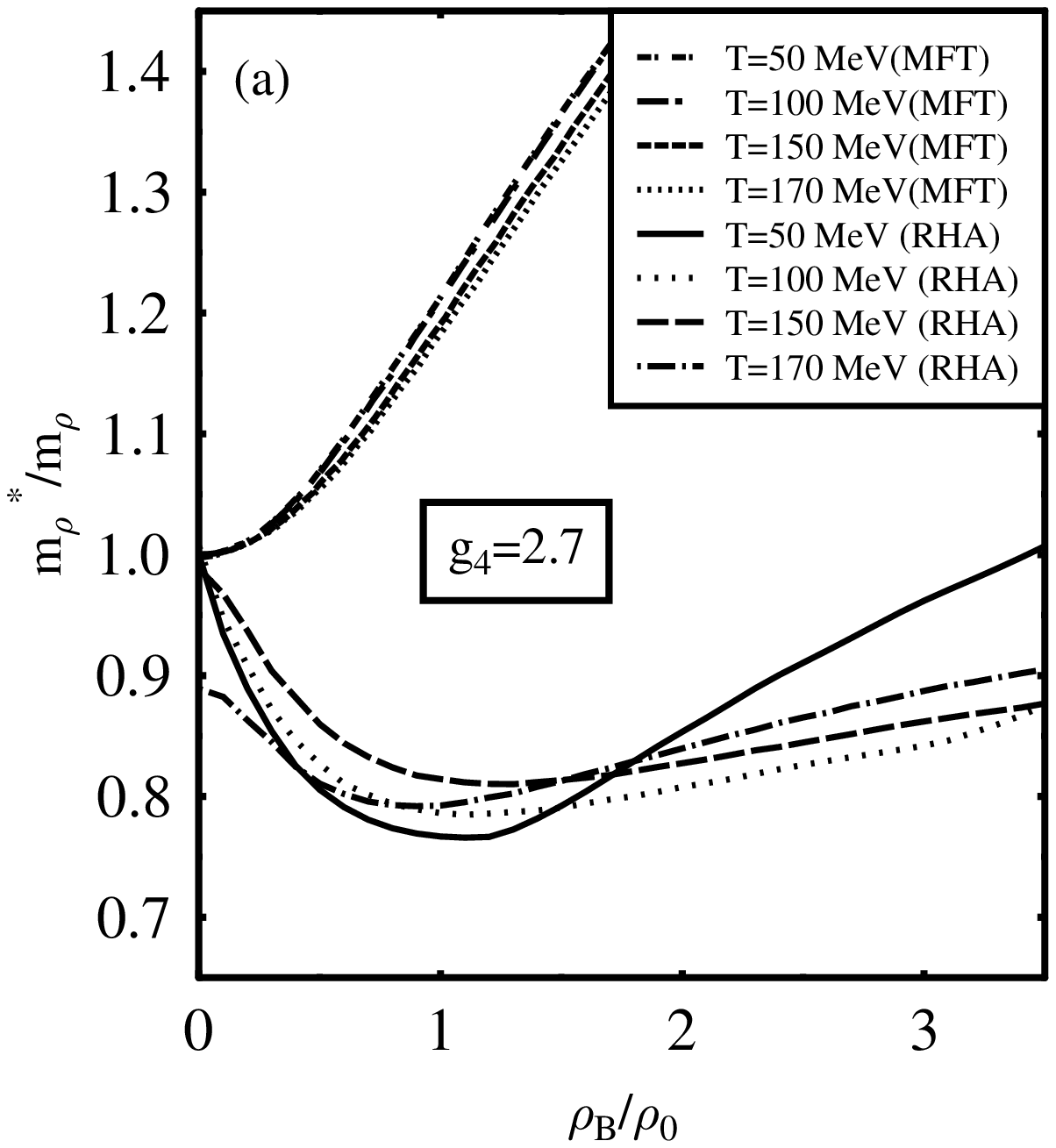}}
\parbox[b]{8cm}{
\includegraphics[width=9.2cm,height=9cm]{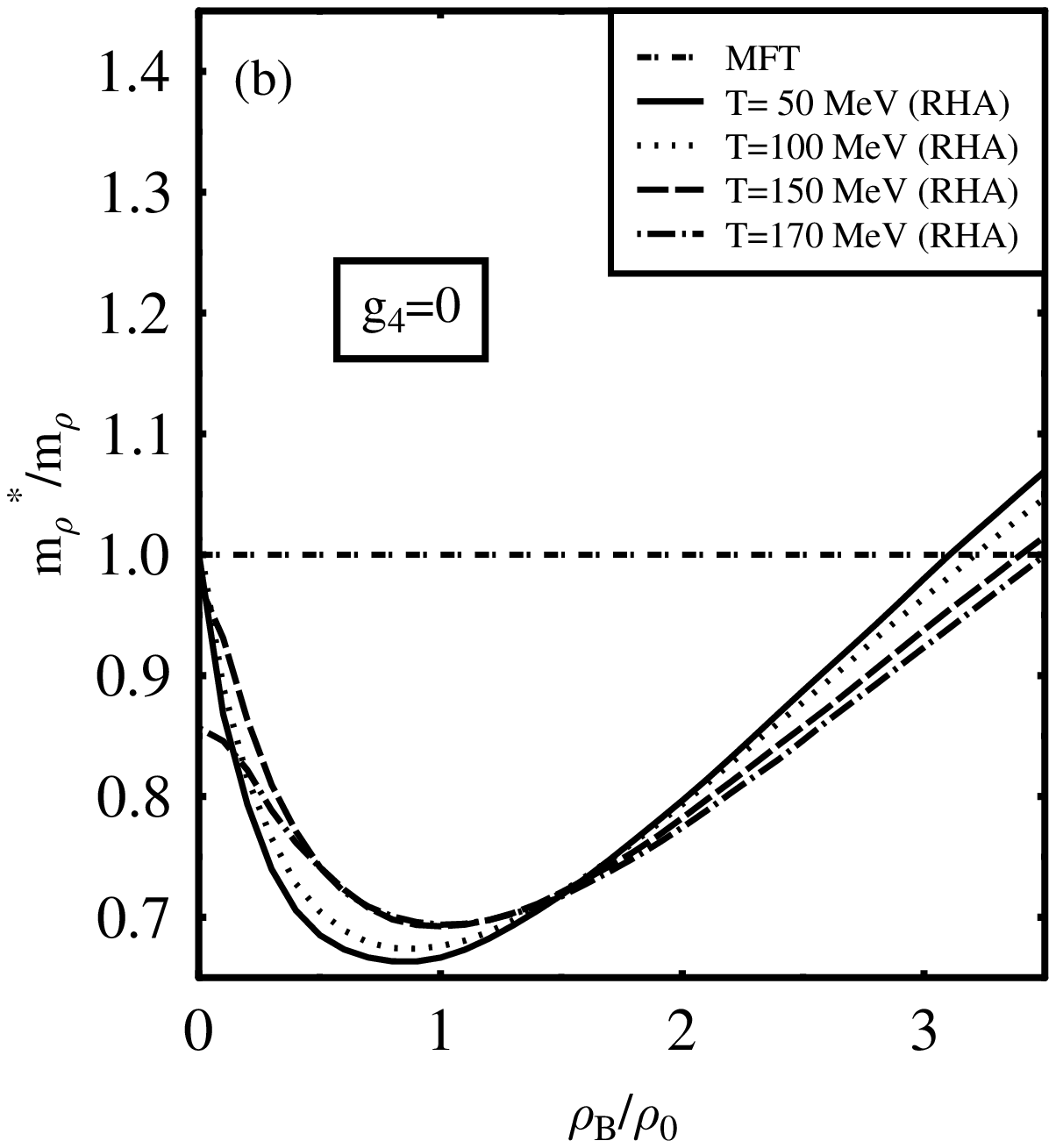}}
\caption{
\label{mrhofig}
Effective $\rho$ mass as a function of density in the mean field
(MFT) and in RHA
for different temperatures and $ f_s = 0 $
for (a) $g_4$=2.7 and (b) $g_4$=0.}
%\end{center}
\end{figure}

The vector meson masses ($\omega$ and $\rho$) as modified by the
interaction to the nucleons in the thermal medium are plotted in
figures \ref{momgfig} and \ref{mrhofig}.
The effective vector meson masses arising due to nucleon-antinucleon
loop in the relativistic Hartree approximation
are compared to the mean field case. 
For the nucleon-rho couplings, the vector and tensor couplings
as obtained from the N-N forward dispersion relation
\cite{hatsuda1,sourav,grein} are used. The medium modified vector
meson masses are plotted in figures \ref{momgfig} and \ref{mrhofig}
with and without the quartic interaction for the vector fields.
The increase of the nucleon masses with temperature
at finite densities is reflected as an increase
in the vector meson masses as was also seen in Ref. \cite{liko}.
If we switch off the quartic self interaction, 
the vector meson masses have no density and
temperature dependence in MFT as seen from (\ref{vmass}).
In RHA, a significant reduction of the $\omega$ and $\rho$ masses
due to the Dirac sea polarization is found up to around nuclear
saturation density. At higher densities, the density dependent
part of the vector meson self energy, describing the interaction
with the Fermi sea fluctuations starts to be more dominating,
leading to increasing masses instead.
But in the case of the $\omega$ mass, above 100 MeV the effect of
the Fermi sea polarization seems not to be sufficient in order
to overturn the original decreasing tendency.
%In contrast, the $\rho$ mass start to rise at $\rho_B/\rho_0\sim 1$,
%supported by the nonvanishing tensor coupling.\\
For $g_4 \neq 0$ the vector meson masses increase monotonically
with density in the mean field case. The mean field value of the
$\omega$ field increases with density and consequently due to 
(\ref{vmass}) the vector meson masses increase. \\
In RHA, the masses drop up to $\rho_0$ but a finite
value for $g_4$ leads to a modified high density behaviour. 
Above, $\rho_0$, the masses increase with density at all temperatures.
It is seen that the density dependence dominates over the temperature
dependence. 

\begin{figure}
%\begin{center}
\parbox[b]{8cm}{
\includegraphics[width=9.2cm,height=9cm]{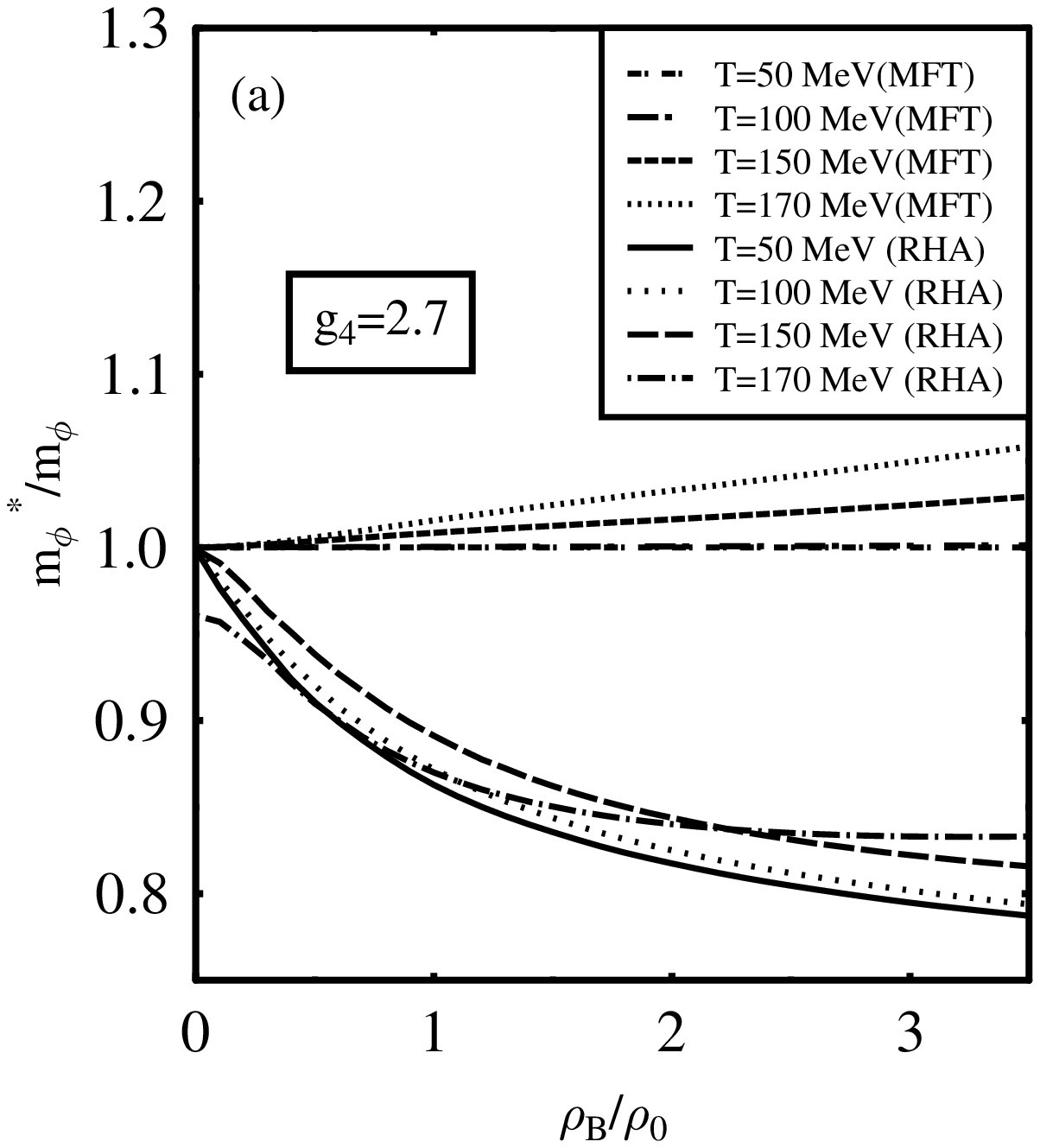}}
\parbox[b]{8cm}{
\includegraphics[width=9.2cm,height=9cm]{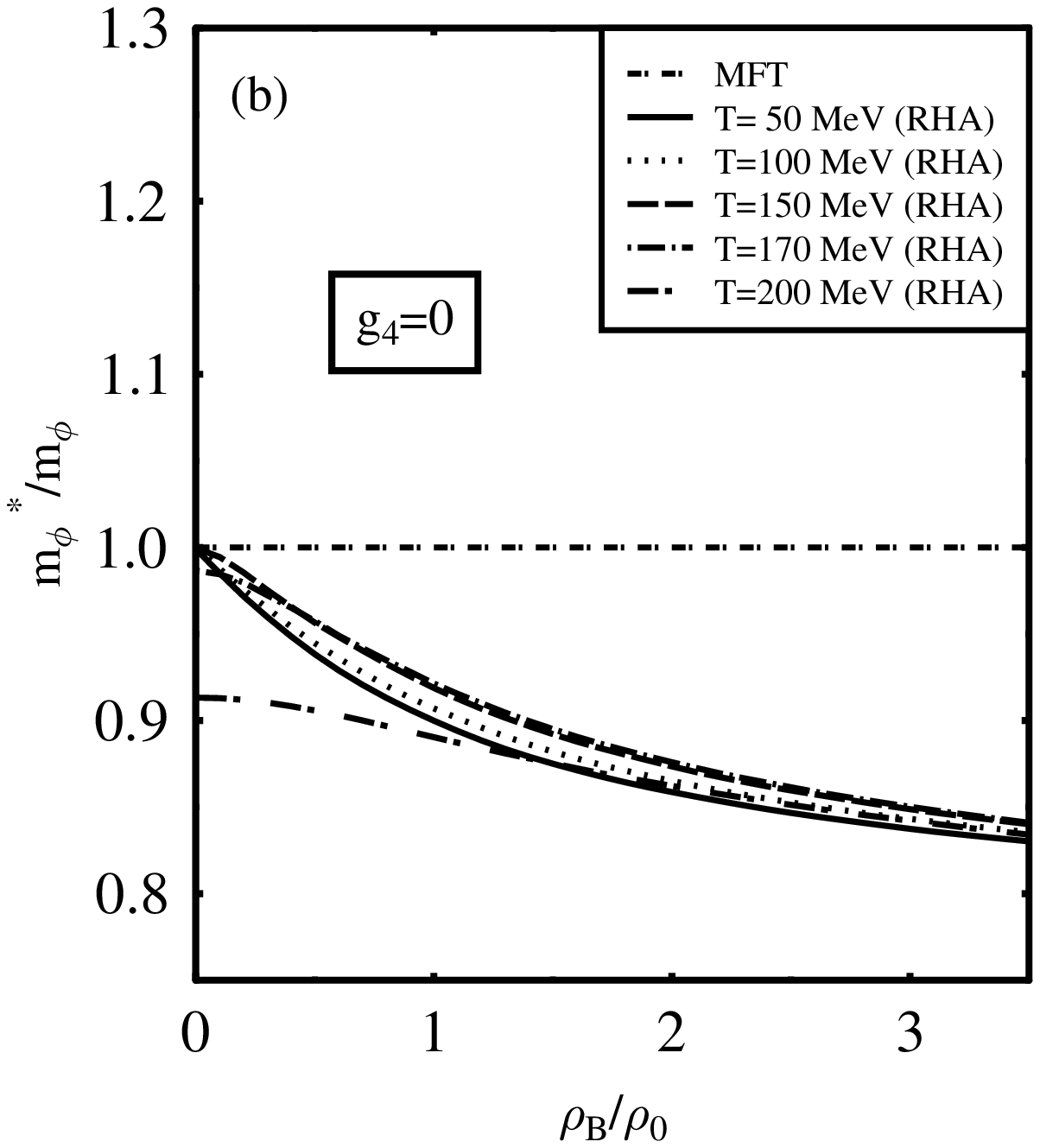}}
\caption{
\label{mphifig}
Effective $\phi$ meson mass in the mean field approximation and
including the Hartree contributions
for (a) $g_4$=2.7 and (b) $g_4$=0.}
\end{figure}

The medium modification for the vector meson $\phi$
in the hyperonic matter is plotted in figure \ref{mphifig}.
One observes that the strange meson $\phi$ has smaller mass modifications
compared to the $\omega$ and $\rho$ mesons.
This is due to the fact that $\phi$ meson does not couple
to the nucleons and also, the hyperon masses are rather insensitive
to the changes in the baryon densities
as may be seen in figures \ref{mlrhtemp} - \ref{mxirhtemp}.
One might note here that unlike
the $\omega$ and $\rho$ vector mesons, where the presence of the
quartic self interaction for the vector fields attenuates
the drop of the meson masses in relativistic Hartree approximation,
the $\phi$-mass has a larger drop when $g_4 \ne 0$.
The reason for this is that the nucleon-$\omega$ coupling
is larger in the presence of this interaction as can be seen
from table I. Hence the hyperon-$\phi$ couplings, which are
related to $g_{N\omega}$ through the relation (\ref{gvb})
are higher in value. This gives rise to a larger drop
of the $\phi$ mass due to the hyperon-antihyperon excitations
in relativistic Hartree approximation. The nonzero $g_4$
gives rise to an increasing contribution to the $\phi$ mass
as proportional to the quadratic strength of the $\phi$-field
unlike the case of $\omega$ and $\rho$ mesons where this
is proportional to the strength of the $\omega$-field.
Since the strength of the $\phi$ field is much smaller
than that of the $\omega$, this increase in the $\phi$-mass
remains small compared to the drop due to the hyperon sea.
%In the presence of strange mesons in the model, though there
%will be contributions to the vector meson masses in the medium,
%through hyperon masses, the qualitative feature of phi-meson undergoing
%smaller medium modification as compared to the omega meson in
%the hyperonic matter is expected to remain unchanged. This is
%because of the fact that the $\omega$ meson gets substantial
%contribution from the coupling to the nucleons, which undergo
%much more medium modification as compared to the hyperons,
%to which the $\phi$ meson does not couple.
The strange meson ($\phi$) mass modification observed  as small
compared to the $\omega$ and $\rho$ meson masses is in line
with the earlier observations \cite{hat,hatsuda1,pal}.
\section{summary}
To summarize, in the present work, we have considered a chiral SU(3) model 
for the description of the hot and strange hadronic matter. The effect of
the baryonic vacuum polarizations has been taken into account
in the relativistic Hartree approximation for study of the hadronic
properties. The coupling of the baryons to
both nonstrange and strange scalar fields modifies the scalar self
energy and is taken into account while summing over the baryonic
tadpole diagrams in the relativistic Hartree approximation.
The vector meson ($\omega$ and $\rho$) masses
are calculated in the thermal medium
arising from the nucleon-antinucleon loop and are seen to have
large drops due to the Dirac sea contribution.
However, the strange vector meson, $\phi$ mass is seen to
be much less modified due to the baryon Dirac sea,
as compared to the nonstrange vector mesons.
The vector meson properties are directly linked to
the dilepton spectra in relativistic heavy ion collision
experiments. It will thus be worth investigating how the dilepton
spectra are modified due to the medium modification of the
vector mesons. The hadron properties in the medium also
modify the number densities and hence would modify the particle
ratios observed in relativistic heavy ion collision
experiments. These have been studied in the mean field approximation
in the present chiral SU(3) model \cite{zsch02} and it will be 
interesting to examine the effects of vacuum polarisations on the 
particle ratios. These and related problems are under investigation.
%
%%%%%%%%%%%%%%%%%%%%%%%%%%%%%%%%%%%%%%%%%%%%%%%%%%%%%%%%%%%%%%%%%%%
\begin{acknowledgements}
%%%%%%%%%%%%%%%%%%%%%%%%%%%%%%%%%%%%%%%%%%%%%%%%%%%%%%%%%%%%%%%%%%%
We thank J. Reinhardt for fruitful discussions.  One of the authors (AM) 
is grateful to the Institut f\"ur Theoretische Physik for warm hospitality 
and acknowledges financial support from Bundesministerium f\"ur Bildung
und Forschung (BMBF). The support from the Frankfurt Center for Scienctific 
Computing (CSC) is gratefully acknowledged. 

\end{acknowledgements}

%%%%%%%%%%%%%%%%%%%%%%%%%%%%%%%%%%%%%%%%%%%%%%%%%%%%%%%%%%%%%%%%%%%%%%%%%%%%
%
%%%%
%

\begin{thebibliography}{1}
%
\bibitem{helios}
 N. Masera for the HELIOS-3 collaboration, Nucl. Phys. {\bf A 590}, 93c (1995).
\bibitem {ceres}
 G. Agakichiev et al (CERES collaboration), Phys. Rev. Lett. {\bf 75}, 1272 (1995);
 G. Agakichiev et al (CERES collaboration), Phys. Lett. {\bf B 422},
 405 (1998); G. Agakichiev et al (CERES collaboration), Nucl. Phys.
 {\bf A 661}, 23c (1999).
\bibitem {dls}
 R. J. Porter et al (DLS collaboration), Phys. Rev.
 Lett. {\bf 79}, 1229 (1997); W. K. Wilson et al (DLS collaboration),
 Phys. Rev. {C 57}, 1865 (1998).
\bibitem {rhic}
 D. P. Morrison (PHENIX collaboration), Nucl. Phys.{\bf A 638}, 565c (1998).
\bibitem {hades}
 J. Stroth (HADES collaboration), Advances in Nucl. Dyn. {\bf 5}, 311 (1999).
\bibitem {brown}
 G. E. Brown and M. Rho, Phys. Rev. Lett. {\bf 66}, 2720 (1991).
\bibitem{rapp}
 R.~Rapp and J.~Wambach, Adv.\ Nucl.\ Phys.\  {\bf 25} (2000) 1;
 R.~Rapp, G.~Chanfray and J.~Wambach, Nucl.\ Phys.\ A {\bf 617} (1997) 472.
\bibitem{hat}
 T. Hatsuda and Su H. Lee, Phys. Rev. {\bf C 46}, R34 (1992);
 T. Hatsuda, Nucl. Phys. {\bf A 544} 27c (1992);
 T. Hatsuda, S. H. Lee and H. Shiomi, Phys. Rev. {\bf C 52}, 3364 (1995).
\bibitem{jin}
 X. Jin and D. B. Leinweber, Phys. Rev. {\bf C 52}, 3344 (1995);
 T. D. Cohen, R. D. Furnstahl, D. K. Griegel and X. Jin,
 Prog. Part. Nucl. Phys. {\bf 35}, 221 (1995); R. Hofmann, Th. Gutsche,
 A. Faessler, Eur. Phys. J. {\bf C 17}, 651 (2000); S. Mallik and
 K. Mukherjee, Phys. Rev. {\bf D 58}, 096011 (1998).
\bibitem {samir}
 S. Mallik and A. Nyffeler, Phys. Rev. {\bf C 63}, 065204 (2001).
\bibitem{weise}
 F. Klingl, N. Kaiser, W. Weise, Nucl. Phys. {\bf A 624},527 (1997).
\bibitem{pisa}
 R. D. Pisarski, Phys. Rev. {\bf D 52}, R3773 (1995);
 Nucl. Phys. {\bf A 590}, 553c (1995).
\bibitem{csong}
 C. Song, Phys. Rev. {\bf D 55}, 3962 (1996).
\bibitem{mharada}
 M. Harada, A. Shibata, Phys. Rev. {\bf D 55}, 6716 (1997).
\bibitem{ernst}
 C. Ernst, S. A. Bass, M. Belkacem, H. St\"ocker
 and W. Greiner, Phys. Rev. {\bf C 58}, 447 (1998).
\bibitem{qhd}
 B. D. Serot and J. D. Walecka, Adv. Nucl. Phys. {\bf 16},
 1 (1986); S. A. Chin, Ann. Phys. (N. Y.) {\bf 108}, 301 (1977).
\bibitem{hatsuda}
 H. Shiomi and T. Hatsuda, Phys. Lett. {\bf B 334}, 281 (1994).
\bibitem {hatsuda1}
 T. Hatsuda, H. Shiomi and H. Kuwabara, Prog. Theor. Phys. {\bf 95},
1009 (1996).
\bibitem{jeans}
 H.-C. Jean, J. Piekarewicz and A. G. Williams, Phys. Rev. {\bf C 49}, 1981 (1994);
 K. Saito, K. Tsushima, A. W. Thomas, A. G. Williams, Phys. Lett. {\bf B 433}, 243 (1998).
\bibitem{sourav}
 Jan-e Alam, S. Sarkar, P. Roy, B. Dutta-Roy and B. Sinha, Phys. Rev. {\bf C 59}, 905 (1999).
\bibitem{caillon}
 J. A. Caillon, J. Labarsouque, Phys. Lett. {\bf B 331} 19 (1993).
\bibitem{deformed}
S. Schramm, Phys. Rev.
{\bf C 66} 064310 (2002).
\bibitem{paper3}
 P. Papazoglou, D. Zschiesche, S. Schramm, J. Schaffner-Bielich, H. St\"ocker,
 and W. Greiner, Phys. Rev. C {\bf 59},  411  (1999).
\bibitem{springer}
 D. Zschiesche, P. Papazoglou, S. Schramm, C. Beckmann, J. Schaffner-Bielich, H.
 St\"ocker, and W. Greiner, Springer Tracts in Modern Physics {\bf 163},  129
 (2000).
\bibitem{hartree}
 D. Zschiesche, A. Mishra, S. Schramm, H. St\"ocker and W. Greiner nucl-th/0302073.
\bibitem{vecmass}
 A. Mishra, J. C. Parikh and W. Greiner, J. Phys. {\bf G 28}, 151 (2002).
\bibitem{dlp}
 A. Mishra, J. Reinhardt, H. St\"ocker and W. Greiner, Phys. Rev. {\bf C 66}, 064902 (2002).
\bibitem{mishra}
 A. Mishra, P. K. Panda, S. Schramm, J. Reinhardt and W. Greiner,
 Phys. Rev. {\bf C 56}, 1380 (1997);
 A. Mishra, P. K. Panda and W. Greiner, J. Phys. {\bf G 27}, 1561 (2001);
 A. Mishra, P. K. Panda and W. Greiner, J. Phys. {\bf G 28}, 67 (2002).
\bibitem{saku69}
 J.~J. Sakurai, {\em Currents and Mesons} (University of Chicago Press, Chicago,
 1969).
\bibitem{gasi69}
S. Gasiorowicz and D. Geffen, Rev. Mod. Phys. {\bf 41},  531  (1969).
\bibitem{mitt68}
P.~K. Mitter and L.~J. Swank, Nucl. Phys. B {\bf 8},  205  (1968).
\bibitem{toki}
Y. Sugahara and H. Toki, Nucl. Phys. {\bf A 579}, 557 (1994).
%\bibitem{sche71}
% J. Schechter and Y. Ueda, Phys. Rev. D {\bf 3},  168  (1971).
\bibitem{sche80}
 J. Schechter, Phys. Rev. D {\bf 21},  3393  (1980).
%\bibitem{serot97}
% B.~D. Serot and J.~D. Walecka, Int. J. Mod. Phys. E {\bf 6},  515  (1997).
\bibitem{kurasawa}
H. Kurasawa, T. Suzuki, Nucl. Phys. {\bf A 490} (1988) 571;
%\bibitem{asakawa}
%M. Asakawa, C. M. Ko, P. Levai and X. J. Qiu,
 %Phys. Rev. {\bf C 46}, R1159 (1992); M. Herrmann, B. L. Friman
 %and W. N\"orenberg, Nucl. Phys. {\bf A 560}, 411 (1993);
 %G. Chanfray and P. Shuck, Nucl. Phys. {\bf A 545}, 271c (1992).
%\bibitem{sakurai}
 %J. J. Sakurai, Currents and Mesons (The University
 %od Chicago Press, Chicago, 1969).
%\bibitem{gellmann}
 %M. Gell-Mann, D. Sharp, and W. D. Wagner, Phys. Rev. Lett. {\bf 8},
% 261 (1962).
%
\bibitem{liko} G. Q. Li, C. M. Ko and G. E. Brown, Nucl. Phys.
{\bf A 606}, 568 (1996).
\bibitem{furnst} R. J. Furnstahl and B. D. Serot, Phys. Rev. {\bf C 41},
 262 (1990).
\bibitem{theis} J. Theis, G. Graebner, G. Buchwald, J. Maruhn,
W. Greiner, J. Polonyi, Phys. Rev. {\bf D 28}, 2286 (1983).
\bibitem{grein}
 W. Grein, Nucl. Phys. {\bf B 131}, 255 (1977);
 W. Grein and P. Kroll, Nucl. Phys. {\bf A 338}, 332 (1980).
\bibitem {pal} S. Pal, Song Gao, H. Stoecker and W. Greiner,
Phys. Lett. {\bf B} 465, 282 (1999).
\bibitem{zsch02} D. Zschiesche, S. Schramm, J. Schaffner-Bielich,
H. St\"ocker, W. Greiner, Phys. Lett. {\bf B 547}, 7 (2002).

\end{thebibliography}
\end{document}